\documentclass[12pt]{article}
\RequirePackage[OT1]{fontenc}
\RequirePackage{amsthm,amsmath}
\RequirePackage[authoryear]{natbib}

\usepackage{amsmath}
\usepackage{graphicx,psfrag,epsf}
\usepackage{enumerate}
\usepackage{natbib}
\usepackage{multirow}

\usepackage{mathrsfs}
\usepackage{amssymb,amsfonts}
\usepackage{bm}
\usepackage{dsfont}
\usepackage{diagbox}
\newtheorem{algorithm}{Algorithm}
\newtheorem{example}{Example}
\newtheorem{lemma}{Lemma}[section]

\newtheorem{remark}{Remark}
\newtheorem{theorem}{Theorem}
\usepackage{xcolor}
\usepackage{epstopdf}
\usepackage[caption=false]{subfig}
\usepackage[latin1]{inputenc}
\usepackage{color}
\usepackage{xcolor}
\usepackage{lineno,hyperref}
\usepackage{multicol}
\usepackage{algorithm}
\usepackage{algpseudocode}
\usepackage{lipsum}
\usepackage{appendix}
\usepackage{amsmath}
\usepackage{graphicx}
\usepackage{lineno}
\usepackage{array}
\usepackage{longtable}
\usepackage{natbib}
\usepackage{float}
\usepackage{comment}
\usepackage{url}
\usepackage{booktabs}
\usepackage{multirow}
\usepackage{natbib}

\begin{document}


\title{\large\bf ForLion: A New Algorithm for D-optimal Designs under General Parametric Statistical Models with Mixed Factors}
\author{Yifei Huang$^1$, Keren Li$^2$,\\ Abhyuday Mandal$^3$, and Jie Yang$^{1}$\\ \\
	$^1$University of Illinois at Chicago,\\ $^2$University of Alabama at Birmingham, and\\ $^3$University of Georgia
}

\maketitle


\begin{abstract}
In this paper, we address the problem of designing an experimental plan with both discrete and continuous factors under fairly general parametric statistical models. We propose a new algorithm, named ForLion, to search for locally optimal approximate designs under the D-criterion. The algorithm performs an exhaustive search in a design space with mixed factors while keeping high efficiency and reducing the number of distinct experimental settings. Its optimality is guaranteed by the general equivalence theorem. We present the relevant theoretical results for multinomial logit models (MLM) and generalized linear models (GLM), and demonstrate the superiority of our algorithm over state-of-the-art design algorithms using real-life experiments under MLM and GLM. Our simulation studies show that the ForLion algorithm could reduce the number of experimental settings by 25\% or improve the relative efficiency of the designs by 17.5\% on average. Our algorithm can help the experimenters reduce the time cost, the usage of experimental devices, and thus the total cost of their experiments while preserving high efficiencies of the designs.
\end{abstract}


\def\thefigure{\arabic{figure}}
\def\thetable{\arabic{table}}

\renewcommand{\theequation}{\arabic{equation}}

\section{Introduction}\label{sec:introduction}

Our research is motivated by an experiment on the emergence of house flies for studying biological controls of disease-transmitting fly species \citep{itepan1995, atkinson1999}. In the original experiment~\citep{itepan1995}, $n=3,500$ pupae were grouped evenly into seven subsets and exposed in a radiation device tuned to seven different gamma radiation levels $x_i \in \{80, 100, 120, 140, 160, 180, 200\}$ in units Gy, respectively.  After a certain period of time, each pupa had one of three possible outcomes: {\tt unopened}, {\tt opened but died} (before completing emergence), or {\tt completed emergence}.  
The total experimental costs in time and expense were closely related to the number of distinct settings of the radiation device. By searching the grid-1 settings in $[80, 200]$ using their lift-one algorithm, \cite{bu2020} proposed a design on five settings $\{80, 122, 123, 157, 158\}$ and improved the relative efficiency of the original design by $20.8\%$ in terms of D-criterion. More recently, \cite{ai2023locally} obtained a design focusing on four settings $\{0, 101.1, 147.8, 149.3\}$ by employing an algorithm that combines the Fedorov-Wynn \citep{fedorov2014} and lift-one \citep{bu2020} algorithms on searching the continuous region $[0, 200]$. Having noticed that both \cite{bu2020}'s and \cite{ai2023locally}'s designs contain pairs of settings that are close to each other, we propose a new algorithm, called the ForLion algorithm (see Section~\ref{sec:general_algorithm}), that incorporates a merging step to combine close experimental settings while maintaining high relative efficiency. For this case, our proposed algorithm identifies a D-optimal design on $\{0, 103.56, 149.26\}$, which may lead to a $40\%$ or $25\%$ reduction of the experimental cost compared to \cite{bu2020}'s or \cite{ai2023locally}'s design, respectively.

In this paper, we consider experimental plans under fairly general statistical models with mixed factors.  The pre-determined design region ${\cal X} \subset \mathbb{R}^d$ with $d\geq 1$ factors is compact, that is, bounded and closed, for typical applications (see Section~2.4 in \cite{fedorov2014}). For many applications, ${\mathcal X} = \prod_{j=1}^d I_j$, where $I_j$ is either a finite set of levels for a qualitative or discrete factor, or an interval $[a_j, b_j]$ for a quantitative or continuous factor. To simplify the notations, we assume that the first $k$ factors are continuous, where $0\leq k\leq d$, and the last $d-k$ factors are discrete.
Suppose $m\geq 2$ distinct experimental settings ${\mathbf x}_1, \ldots, {\mathbf x}_m \in {\cal X}$, known as the {\it design points}, are chosen, and $n > 0$ experimental units are available for the experiment with $n_i \geq 0$ subjects allocated to ${\mathbf x}_i$, such that, $n=\sum_{i=1}^m n_i$~. We assume that the responses, which could be vectors, are independent and follow a parametric model $M({\mathbf x}_i; \boldsymbol\theta)$ with some unknown parameter(s) $\boldsymbol\theta \in \mathbb{R}^p$, $p\geq 1$. 
In the design theory, ${\mathbf w} = (w_1, \ldots, w_m)^T = (n_1/n, \ldots, n_m/n)^T$, known as an {\it approximate allocation}, is often considered instead of the {\it exact allocation} ${\mathbf n} = (n_1, \ldots, n_m)^T$ (see, for examples, \cite{kiefer1974}, Section~1.27 in \cite{pukelsheim2006optimal}, and Section~9.1 in \cite{atkinson2007}). Under regularity conditions, the corresponding Fisher information matrix is ${\mathbf F} = \sum_{i=1}^m w_i{\mathbf F}_{{\mathbf x}_i} \in \mathbb{R}^{p\times p}$ up to a constant $n$, where ${\mathbf F}_{{\mathbf x}_i}$ is the Fisher information at ${\mathbf x}_i$~.
In this paper, the design under consideration takes the form of $\boldsymbol\xi = \{({\mathbf x}_i, w_i), i=1, \ldots, m\}$, where $m$ is a flexible positive integer, ${\mathbf x}_1, \ldots, {\mathbf x}_m$ are distinct design points from ${\cal X}$, $0\leq w_i \leq 1$, and $\sum_{i=1}^m w_i = 1$. 
We also denote $\boldsymbol\Xi = \{\{({\mathbf x}_i, w_i), i=1, \ldots, m\} \mid m\geq 1; {\mathbf x}_i \in {\cal X}, 0\leq w_i\leq 1, i=1, \ldots, m; \sum_{i=1}^m w_i = 1\}$ be the collection of all feasible designs. 

Under different criteria, such as D-, A-, or E-criterion (see, for example, Chapter 10 in  \cite{atkinson2007}), many numerical algorithms have been proposed for finding optimal designs. If all factors are discrete, the design region ${\mathcal X}$ typically contains a finite number of design points, still denoted by $m$. Then the design problem is to optimize the approximate allocation ${\mathbf w} = (w_1, \ldots, w_m)^T$. Commonly used design algorithms include Fedorov-Wynn \citep{fedorov1972, fedorov1997model}, multiplicative \citep{titterington1976, titterington1978, silvey1978}, cocktail \citep{yu2011d}, and lift-one \citep{ym2015, ytm2016, bu2020}, etc. Besides, classical optimization techniques such as Nelder-Mead \citep{nelder1965simplex}, quasi-Newton \citep{broyden1965class, dennis1977quasi}, conjugate gradient \citep{hestenes1952methods, fletcher1964function}, and simulated annealing \citep{kirkpatrick1983optimization} may also be used for the same purpose \citep{nocedal2006}. A comprehensive numerical study by \cite{ymm2016} (Table~2) showed that the lift-one algorithm outperforms commonly used optimization algorithms in identifying optimal designs, resulting in designs with fewer points.  

Furthermore, many deterministic optimization methods may also be used for finding optimal designs under similar circumstances. Among them, polynomial time (P-time) methods including linear programming  \citep{HarmanRadoslav2008Cc-e}, second-order cone programming \citep{SagnolGuillaume2011Codo}, semidefinite programming \citep{DuarteBelmiroP.M.2018Aabo, DuarteBelmiroP.M.2015FBOD, venables2002modern, WongWengKee2023UCtC,YeJaneJ2013MtCN}, mixed integer linear programming \citep{Vo-ThanhNha2018Sbim}, mixed integer quadratic programming \citep{HarmanRadoslav2014Ceed}, mixed integer second-order cone programming \citep{SagnolGuillaume2015CEDD}, and mixed integer semidefinite programming \citep{DuarteBelmiroP.M.2023EODo}, are advantageous for discrete grids due to their polynomial time complexity and capability of managing millions of constraints efficiently. Notably, nonlinear polynomial time (NP-time) methods, such as nonlinear programming \citep{DuarteBelmiroP.M.2022ODoE}, semi-infinite programming \citep{DuarteBelmiroP.M.2014Aspb}, and mixed integer nonlinear programming \citep{DuarteBelmiroP.M.2020Oedo} have been utilized as well. 

When the factors are continuous, the Fedorov-Wynn algorithm can still be used by adding a new design point in each iteration, which maximizes a sensitivity function on ${\mathcal X}$  \citep{fedorov2014}. To improve the efficiency, \cite{ai2023locally} proposed a new algorithm for D-optimal designs under a continuation-ratio link model with continuous factors, which essentially incorporates the Fedorov-Wynn (for adding new design points) and lift-one (for optimizing the approximate allocation) algorithms. Nevertheless, the Fedorov-Wynn step tends to add unnecessary closely-spaced design points (see Section~\ref{sec:MLM_example_fly}), which may increase the experimental cost. An alternative approach is to discretize the continuous factors and consider only the grid points \citep{yangmin2013}, which may be computationally expensive especially when the number of factors is moderate or large.

Little has been done to construct efficient designs with mixed factors. \cite{lukemire2018} proposed the $d$-QPSO algorithm,  a modified quantum-behaved particle swarm optimization (PSO) algorithm, for D-optimal designs under generalized linear models with binary responses. 
Later, \cite{lukemire2022} extended the PSO algorithm for locally D-optimal designs under the cumulative logit model with ordinal responses. However, like other stochastic optimization algorithms, the PSO-type algorithms cannot guarantee that an optimal solution will ever be found \citep{kennedy1995particle, poli2007particle}.

Following \cite{ai2023locally} and \cite{lukemire2018, lukemire2022}, we choose D-criterion, which maximizes the objective function $f({\boldsymbol\xi}) = \left|\mathbf F(\boldsymbol \xi) \right| = \left|\sum_{i=1}^m w_i {\mathbf F}_{{\mathbf x}_i}\right|$, $\boldsymbol\xi \in \boldsymbol\Xi$. Throughout this paper, we assume $f({\boldsymbol\xi}) > 0$ for some $\boldsymbol\xi \in \boldsymbol\Xi$ to avoid trivial optimization problems. Unlike \cite{bu2020} and \cite{ai2023locally}, the proposed ForLion algorithm does not need to assume ${\rm rank}({\mathbf F}_{{\mathbf x}}) < p$ for all ${\mathbf x} \in {\cal X}$ (see Remark~\ref{rem:lift-one_step} and Example~\ref{ex:rank_F_p}). Compared with the PSO-type algorithms for similar purposes \citep{lukemire2018, lukemire2022}, our ForLion algorithm could improve the relative efficiency of the designs significantly (see Example~\ref{ex:ESD} for an electrostatic discharge experiment discussed by \cite{lukemire2018} and Section~\ref{sec:MLM_example_surface} of the Supplementary Material for a surface defects experiment \citep{phadke1989, wu2008, lukemire2022}). 
Our strategies may be extended to other optimality criteria, such as, A-optimality, which minimizes the trace of the inverse of the Fisher information matrix,
and E-optimality, which maximizes the smallest eigenvalue of the Fisher information matrix
(see, for example, \cite{atkinson2007}). 

The remaining parts of this paper are organized as follows. In Section~\ref{sec:general_algorithm}, we present the ForLion algorithm for general parametric statistical models. In Section~\ref{sec:MLM}, we derive the theoretical results for multinomial logistic models (MLM) and revisit the motivated example to demonstrate our algorithm's performance with mixed factors under general parametric models. In Section~\ref{sec:GLM}, we specialize our algorithm for generalized linear models (GLM) to enhance computational efficiency by using model-specified formulae and iterations. We use simulation studies to show the advantages of our algorithm. 
We conclude in Section~\ref{sec:conclusion}.

\section{ForLion for D-optimal Designs with Mixed Factors} \label{sec:general_algorithm}

In this section, we propose a new algorithm, called the {\it ForLion} (\underline{F}irst-\underline{or}der \underline{Li}ft-\underline{on}e) algorithm, for constructing locally D-optimal approximate designs under a general parametric model $M({\mathbf x}; $ $\boldsymbol{\theta})$ with ${\mathbf x} \in {\mathcal X} \subset \mathbb{R}^d$, $d\geq 1$ and $\boldsymbol\theta \in \mathbb{R}^p$, $p\geq 1$. As mentioned earlier, the design region ${\mathcal X} = \prod_{j=1}^d I_j$, where $I_j = [a_j, b_j]$ for $1\leq j\leq k$, $-\infty < a_j < b_j < \infty$, and $I_j$ is a finite set of at least two distinct numerical levels for $j>k$. To simplify the notation, we still denote $a_j = \min I_j$ and $b_j = \max I_j$ even if $I_j$ is a finite set. 

In this paper, we assume $1\leq k\leq d$. That is, there is at least one continuous factor. For cases with $k=0$, that is, all factors are discrete, one may use the lift-one algorithm for general parametric models (see Remark~\ref{rem:lift-one_step}). The goal in this study is to find a design $\boldsymbol\xi = \{({\mathbf x}_i, w_i), i=1, \ldots, m\} \in \boldsymbol{\Xi}$ maximizing $f({\boldsymbol\xi}) = \left|{\mathbf F}(\boldsymbol\xi)\right|$, the determinant of ${\mathbf F}(\boldsymbol\xi)$, where ${\mathbf F}(\boldsymbol\xi) = \sum_{i=1}^m w_i {\mathbf F}_{{\mathbf x}_i} \in \mathbb{R}^{p\times p}$. Here $m\geq 1$ is flexible. 

\footnotesize
\begin{algorithm}
\caption{\bf ForLion} \label{algo:ForLion_general}
\begin{itemize}
\item[$0^\circ$] Set up tuning parameters $\delta > 0$ as the merging threshold and $\epsilon > 0$ as the converging threshold. For example, $\delta = 10^{-6}$ and $\epsilon = 10^{-12}$.

\item[$1^\circ$] Construct an initial design ${\boldsymbol\xi}_0 = \{({\mathbf x}_i^{(0)}, w_i^{(0)}), i=1, \ldots, m_0\}$ such that (i) $\|{\mathbf x}_i^{(0)} - {\mathbf x}_j^{(0)}\| \geq \delta$ for any $i\neq j$; and (ii) $|{\mathbf F}({\boldsymbol\xi}_0)| > 0$.
For example, one may sequentially and randomly choose ${\mathbf x}_i^{(0)}$ from either $\prod_{j=1}^d \{a_j, b_j\}$ or $\prod_{j=1}^d I_j$ such that the new point is at least $\delta$ away from the previous points, until some $m_0$ such that $|\sum_{i=1}^{m_0} {\mathbf F}_{{\mathbf x}_i^{(0)}}| > 0$. The weights $w_i^{(0)}$ may be defined uniformly (all equal to $1/m_0$) or randomly (proportional to $U_i$ with $U_i$'s i.i.d.~from an exponential distribution). 

\item[$2^\circ$] Merging step: Given the design ${\boldsymbol\xi}_t = \{({\mathbf x}_i^{(t)}, w_i^{(t)}), i=1, \ldots, m_t\}$ at the $t$th iteration, check the pairwise Euclidean distances among ${\mathbf x}_i^{(t)}$'s. If there exist $1\leq i < j\leq m_t$, such that, $\|{\mathbf x}_i^{(t)} - {\mathbf x}_j^{(t)}\| < \delta$, then merge the two points into a new point $({\mathbf x}_i^{(t)} + {\mathbf x}_j^{(t)})/2$ with weight $w_i^{(t)} + w_j^{(t)}$, and replace $m_t$ by $m_t-1$. Repeat the procedure till any two remaining points have a distance of at least $\delta$.

\item[$3^\circ$] Lift-one step: Given ${\boldsymbol\xi}_t$, run the lift-one algorithm (see Remark~\ref{rem:lift-one_step}) with converging threshold $\epsilon$ to find the converged allocation $w_1^*, \ldots, w_{m_t}^*$ for the design points ${\mathbf x}_1^{(t)}, \ldots, {\mathbf x}_{m_t}^{(t)}$. Replace $w_i^{(t)}$'s with $w_i^*$'s, respectively.

\item[$4^\circ$] Deleting step: Update $\boldsymbol\xi_t$ by removing  all ${\mathbf x}_i^{(t)}$'s that have $w_i^{(t)}=0$.

\item[$5^\circ$] New point step: Given ${\boldsymbol\xi}_t$, find a point ${\mathbf x}^* = (x_1^*, \ldots, x_d^*)^T \in {\mathcal X}$ that maximizes $d({\mathbf x}, {\boldsymbol\xi}_t) = {\rm tr}({\mathbf F}({\boldsymbol\xi}_t)^{-1} {\mathbf F}_{\mathbf x})$.
Recall that the first $k$ factors are continuous. If $1\leq k < d$, we denote ${\mathbf x}_{(1)} = (x_1, \ldots, x_k)^T$ and ${\mathbf x}_{(2)} = (x_{k+1}, \ldots, x_d)^T$. Then ${\mathbf x} = ({\mathbf x}_{(1)}^T, {\mathbf x}_{(2)}^T)^T$. Fixing each ${\mathbf x}_{(2)} \in \prod_{j=k+1}^d I_j$, we use the ``L-BFGS-B"  quasi-Newton method \citep{byrd1995} to find  
\[
{\mathbf x}_{(1)}^* = {\rm argmax}_{{\mathbf x}_{(1)} \in \prod_{i=1}^k [a_i, b_i]} d(({\mathbf x}_{(1)}^T, {\mathbf x}_{(2)}^T)^T, {\boldsymbol\xi}_t) 
\]
Note that ${\mathbf x}_{(1)}^*$ depends on ${\mathbf x}_{(2)}$.
Then ${\mathbf x}^*$ is obtained by finding the ${\mathbf x}_{(2)}^*$ associated with the largest $d((({\mathbf x}^*_{(1)})^T, {\mathbf x}_{(2)}^T)^T, {\boldsymbol\xi}_t)$. 
If $k=d$, that is, all factors are continuous, we can always find ${\mathbf x}^* = {\rm argmax}_{{\mathbf x}\in {\mathcal X}} d({\mathbf x}, \boldsymbol{\xi}_t)$ directly.

\item[$6^\circ$] If $d({\mathbf x}^*, {\boldsymbol\xi}_t) \leq p$, go to Step~$7^\circ$. Otherwise, we let ${\boldsymbol\xi}_{t+1} = {\boldsymbol\xi}_t \bigcup \{({\mathbf x}^*, 0)\}$, replace $t$ by $t+1$, and go back to Step~$2^\circ$.

\item[$7^\circ$] Report ${\boldsymbol\xi}_t$ as the D-optimal design.
\end{itemize}
\end{algorithm}
\normalsize

Given a design $\boldsymbol\xi = \{({\mathbf x}_i, w_i), i=1, \ldots, m\}$ reported by the ForLion algorithm (see Algorithm~\ref{algo:ForLion_general}), the general equivalence theorem \citep{kiefer1974, pukelsheim1993, stufken2012, fedorov2014} guarantees its D-optimality on ${\mathcal X}$. As a direct conclusion of Theorem~2.2 in \cite{fedorov2014}, we have the theorem as follows under the regularity conditions (see Section~\ref{subsec:assumptions} in the Supplementary Material, as well as Assumptions (A1), (A2) and (B1) $\sim$ (B4) in Section~2.4 of \cite{fedorov2014}).

\begin{theorem}\label{thm:doptimality}
Under regularity conditions, there exists a D-optimal design that contains no more than $p(p+1)/2$ design points.   
Furthermore, if $\boldsymbol{\xi}$ is obtained by Algorithm~\ref{algo:ForLion_general}, it must be D-optimal. 
\end{theorem}

We relegate the proof of Theorem~\ref{thm:doptimality} and others to Section~\ref{subsec:proofs} of the Supplementary Material.

\begin{remark}\label{rem:lift-one_step}
{\bf Lift-one step for general parametric models:}\quad {\rm
For commonly used parametric models, ${\rm rank}({\mathbf F}_{\mathbf x}) < p$ for each ${\mathbf x} \in {\mathcal X}$. For example, all GLMs satisfy ${\rm rank}({\mathbf F}_{\mathbf x}) = 1$ (see Section~\ref{sec:GLM}).
However, there exist special cases that ${\rm rank}({\mathbf F}_{\mathbf x}) = p$ for almost all ${\mathbf x} \in {\mathcal X}$ (see Example~\ref{ex:rank_F_p} in Section~\ref{sec:F_x_for_MLM}). 

The original lift-one algorithm (see Algorithm~3 in the Supplementary Material of \cite{huang2023constrained}) requires $0\leq w_i < 1$ for all $i=1, \ldots, m$, given the current allocation ${\mathbf w} = (w_1, \ldots, w_m)^T$. If ${\rm rank}({\mathbf F}_{{\mathbf x}_i}) < p$ for all $i$, then $f(\boldsymbol{\xi}) > 0$ implies $0\leq w_i <1$ for all $i$. In that case, same as in the original lift-one algorithm, we define the allocation function as
\[
{\mathbf w}_i(z) = \left(\frac{1-z}{1-w_i} w_1, \ldots, \frac{1-z}{1-w_i} w_{i-1}, z,  \frac{1-z}{1-w_i} w_{i+1}, \ldots, \frac{1-z}{1-w_i} w_m\right)^T = \frac{1-z}{1-w_i} {\mathbf w} + \frac{z-w_i}{1-w_i} {\mathbf e}_i 
\]
where ${\mathbf e}_i = (0, \ldots, 0, 1, 0, \ldots, 0)^T \in \mathbb{R}^m$, whose $i$th coordinate is $1$, and $z$ is a real number in $[0,1]$, such that $\mathbf w_i(z) = \mathbf w_i$ at $z=w_i$, and $\mathbf w_i(z)=\mathbf e_i$ at $z=1$. However, if ${\rm rank}({\mathbf F}_{{\mathbf x}_i}) = p$ and $w_i=1$ for some $i$, we still have $f(\boldsymbol{\xi}) > 0$, but the above ${\mathbf w}_i(z)$ is not well defined. In that case, we define the allocation function in the ForLion algorithm as
\[
{\mathbf w}_i(z) = \left(\frac{1-z}{m-1}, \ldots, \frac{1-z}{m-1}, z, \frac{1-z}{m-1}, \ldots, \frac{1-z}{m-1}\right)^T = \frac{m(1-z)}{m-1}{\mathbf w}_u + \frac{mz-1}{m-1}{\mathbf e}_i
\]
where ${\mathbf w}_u = (1/m, \ldots, 1/m)^T \in \mathbb{R}^m$ is a uniform allocation. For $j\neq i$, we define ${\mathbf w}_j(z) = (1-z) {\mathbf e}_i + z {\mathbf e}_j$~. The rest parts are the same as the original life-one algorithm.
}\hfill{$\Box$}
\end{remark}

\begin{remark}\label{rem:stopping_rule}
{\bf Convergence in finite iterations:}\quad {\rm
In practice, we may relax the stopping rule $d({\mathbf x}^*, {\boldsymbol\xi}_t) \leq p$ in Step~$6^\circ$ of Algorithm~\ref{algo:ForLion_general} to $d({\mathbf x}^*, {\boldsymbol\xi}_t) \leq p + \epsilon$, where $\epsilon$ could be the same as in Step~$0^\circ$. By Section~2.5 in \cite{fedorov2014}, $f((1-\alpha){\boldsymbol\xi}_t + \alpha {\mathbf x}^*) - f({\boldsymbol\xi}_t) \approx \alpha(d({\mathbf x}^*, {\boldsymbol\xi}_t) - p)$ for small enough $\alpha > 0$, where $(1-\alpha){\boldsymbol\xi}_t + \alpha {\mathbf x}^*$ is the design $\{({\mathbf x}_i^{(t)}, (1-\alpha) w_i^{(t)}), i=1, \ldots, m_t\} \bigcup \{({\mathbf x}^*, \alpha)\}$. Thus, if we find an ${\mathbf x}^*$, such that, $d({\mathbf x}^*, {\boldsymbol\xi}_t) > p + \epsilon$, then there exists an $\alpha_0 \in (0,1)$, such that, $f((1-\alpha_0){\boldsymbol\xi}_t + \alpha_0 {\mathbf x}^*) - f({\boldsymbol\xi}_t) > \alpha_0(d({\mathbf x}^*, {\boldsymbol\xi}_t) - p)/2 > \alpha_0 \epsilon/2$. For small enough merging threshold $\delta$ (see Steps~$0^\circ$ and $2^\circ$), we can still guarantee that $f({\boldsymbol\xi}_{t+1}) - f({\boldsymbol\xi}_t) > \alpha_0\epsilon/4$ after Step~$2^\circ$. Under regularity conditions, ${\mathcal X}$ is compact, and $f(\boldsymbol{\xi})$ is continuous and bounded. Our algorithm is guaranteed to stop in finite steps. Actually, due to the lift-one step (Step~$3^\circ$), $f({\boldsymbol\xi}_t)$ is improved fast, especially in the first few steps. For all the examples explored in this paper, our algorithm works efficiently.
}\hfill{$\Box$}
\end{remark}

\begin{remark}\label{rem:distance}
{\bf Distance among design points:}\quad {\rm
In Step~$1^\circ$ of Algorithm~\ref{algo:ForLion_general}, an initial design is selected such that $\|{\mathbf x}_i^{(0)} - {\mathbf x}_j^{(0)}\| \geq \delta$, and in Step~$2^\circ$, two design points are merged if $\|{\mathbf x}_i^{(t)} - {\mathbf x}_j^{(t)}\| < \delta$. The algorithm uses the Euclidean distance as a default metric. Nevertheless, to take the effects of ranges or units across factors into consideration, one may choose a different distance, for example, a normalized distance, such that, $\|{\mathbf x}_i - {\mathbf x}_j\|^2 = \sum_{l=1}^d \left(\frac{x_{il}-x_{jl}}{b_l-a_l}\right)^2$, where ${\mathbf x}_i = (x_{i1}, \ldots, x_{id})^T$ and ${\mathbf x}_j = (x_{j1}, \ldots, x_{jd})^T$. Another useful distance is to define $\|{\mathbf x}_i - {\mathbf x}_j\| = \infty$ whenever their discrete factor levels are different, that is, $x_{il}\neq x_{jl}$ for some $l>k$. Such a distance does not allow any two design points that have distinct discrete factor levels to merge, which makes a difference when $\delta$ is chosen to be larger than the smallest difference between discrete factor levels.
Note that the choice of distance and $\delta$ (see Section~\ref{sec:mergedistance_delta} in the Supplementary Material) won't affect the continuous search for a new design point in Step~$5^\circ$. It is different from the adaptive grid strategies used in the literature \citep{DuarteBelmiroP.M.2018Agsp, harman2020randomized, harman2020greedy} for optimal designs, where the grid can be increasingly reduced in size to locate the support points more accurately.
}\hfill{$\Box$}
\end{remark}

\begin{remark}\label{rem:global_maxima}
{\bf Global maxima:}\quad {\rm
According to Theorem~\ref{thm:doptimality}, when Algorithm~\ref{algo:ForLion_general} converges, that is, $\max_{{\mathbf x} \in {\mathcal X}} d({\mathbf x}, {\boldsymbol\xi}_t) \leq p$, the design ${\boldsymbol\xi}_t$ must be D-optimal. Nevertheless, from a practical point of view, there are two possible issues that may occur. Firstly, the algorithm may fail to find the global maxima in Step~$5^\circ$, which may happen even with the best optimization software \citep{givens2013}. As a common practice (see, for example, Section~3.2 in \cite{givens2013}), one may randomly generate multiple (such as 3 or 5) starting points when finding ${\mathbf x}_{(1)}^*$ in Step~$5^\circ$, and utilize the best one among them. Secondly,  the global maxima or D-optimal design may not be unique (see, for example, Remark~2 in \cite{ymm2016}). In that case, one may keep a collection of D-optimal designs that the algorithm can find. Due to the log-concavity of the D-criterion (see, for example, \cite{fedorov1972}), any convex combinations of D-optimal designs are still D-optimal.
}\hfill{$\Box$}
\end{remark}

\section{D-optimal Designs for MLMs}\label{sec:MLM}

In this section, we consider experiments with categorical responses. Following \cite{bu2020}, given $n_i > 0$ experimental units assigned to a design setting ${\mathbf x}_i \in {\mathcal X}$, the summarized responses ${\mathbf Y}_i = (Y_{i1}, \ldots, Y_{iJ})^T$ follow ${\rm Multinomial}(n_i; \pi_{i1}, \ldots, \pi_{iJ})$ with categorical probabilities $\pi_{i1}, \ldots, \pi_{iJ}$, where $J\geq 2$ is the number of categories, and $Y_{ij}$ is the number of experimental units with responses in the $j$th category. Multinomial logistic models (MLM) have been commonly used for modeling categorical responses \citep{pmcc1995, atkinson1999, bu2020}. A general MLM can be written as 
 \begin{equation}\label{logitunifiedmodel}
  {\mathbf C}^T\log({\mathbf L}{\boldsymbol\pi}_i)={\boldsymbol\eta}_i={\mathbf X}_i{\boldsymbol\theta}, \qquad     i=1,\cdots,m
 \end{equation}
where ${\boldsymbol\pi}_i = (\pi_{i1}, \ldots, \pi_{iJ})^T$ satisfying $\sum_{j=1}^J \pi_{ij}=1$, ${\boldsymbol\eta}_i = (\eta_{i1}, \ldots, \eta_{iJ})^T$, $\mathbf C$ is a $(2J-1) \times J$ constant matrix, $\mathbf X_i$ is the model matrix of $J \times p$ at ${\mathbf x}_i$, $\boldsymbol\theta \in \mathbb{R}^p$ are model parameters, and $\mathbf L$ is a $(2J-1) \times J$ constant matrix taking different forms for four special classes of MLM models, namely, baseline-category, cumulative, adjacent-categories, and continuation-ratio logit models (see \cite{bu2020} for more details). When $J=2$, all the four logit models are essentially logistic regression models for binary responses, which belong to generalized linear models (see Section~\ref{sec:GLM}).

\subsection{Fisher information ${\mathbf F}_{{\mathbf x}}$ and sensitivity function \texorpdfstring{$d({\mathbf x}, \boldsymbol\xi)$}{Lg}}\label{sec:F_x_for_MLM}

The $p\times p$ matrix ${\mathbf F}_{{\mathbf x}}$, known as the Fisher information at ${\mathbf x} \in {\cal X}$, plays a key role in the ForLion algorithm. We provide its formula in detail in Theorem~\ref{thm:F_x}.

\medskip
\begin{theorem}\label{thm:F_x}
For MLM~\eqref{logitunifiedmodel}, the Fisher information ${\mathbf F}_{\mathbf x}$ at ${\mathbf x} \in {\cal X}$ (or ${\cal X}_{\boldsymbol\theta}$ for cumulative logit models, see \cite{bu2020}) can be written as a block matrix  $({\mathbf F}^{\mathbf x}_{st})_{J\times J} \in \mathbb{R}^{p\times p}$, where ${\mathbf F}^{\mathbf x}_{st}$, a sub-matrix of ${\mathbf F}_{\mathbf x}$ with block row index $s$ and column index $t$ in $\{1, \ldots, J\}$, is given by
{\footnotesize
\[
\left\{\begin{array}{cl}
u_{st}^{\mathbf x} \cdot {\mathbf h}_s({\mathbf x}) {\mathbf h}_t({\mathbf x})^T & \mbox{, for }1\leq s, t\leq J-1\\
\sum_{j=1}^{J-1} u_{jt}^{\mathbf x} \cdot {\mathbf h}_c({\mathbf x}) {\mathbf h}_t({\mathbf x})^T & \mbox{, for }s=J, 1\leq t\leq J-1\\
\sum_{j=1}^{J-1} u_{sj}^{\mathbf x} \cdot {\mathbf h}_s({\mathbf x}) {\mathbf h}_c({\mathbf x})^T & \mbox{, for }1\leq s\leq J-1, t=J\\
\sum_{i=1}^{J-1} \sum_{j=1}^{J-1} u_{ij}^{\mathbf x} \cdot {\mathbf h}_c({\mathbf x}) {\mathbf h}_c({\mathbf x})^T & \mbox{, for }s=J, t=J
\end{array}\right.
\]}
where ${\mathbf h}_j({\mathbf x})$ and ${\mathbf h}_c({\mathbf x})$ are predictors at ${\mathbf x}$, and  $u_{st}^{\mathbf x}$'s are known functions of ${\mathbf x}$ and $\boldsymbol{\theta}$ (more details can be found in Appendix~\ref{sec:computing_F_x}).
\end{theorem}

In response to Remark~\ref{rem:lift-one_step} in Section~\ref{sec:general_algorithm}, we provide below a surprising example that the Fisher information ${\mathbf F}_{\mathbf x}$ at a single point ${\mathbf x}$ is positive definite for almost all ${\mathbf x} \in {\cal X}$.

\medskip
\begin{example}{\bf Positive Definite ${\mathbf F}_{\mathbf x}$}\label{ex:rank_F_p}\quad {\rm
We consider a special MLM~\eqref{logitunifiedmodel} with non-proportional odds (npo) (see Section~S.8 in the Supplementary Material of \cite{bu2020} for more details). Suppose $d=1$ and a feasible design point ${\mathbf x} = x \in [a, b] = {\cal X}$, $J\geq 3$, ${\mathbf h}_1({\mathbf x}) = \cdots = {\mathbf h}_{J-1}({\mathbf x}) \equiv x$. The model matrix at ${\mathbf x} = x$ is 
\footnotesize
\[
{\mathbf X}_x = \left(
\begin{array}{cccc}
x & 0 & \cdots & 0\\
0 & x & \cdots & 0\\
\vdots & \vdots & \ddots & \vdots\\
0 & 0 & \cdots & x\\
0 & 0 & \cdots & 0
\end{array}\right)_{J\times (J-1)}
\]
\normalsize
with $p=J-1$. That is, the model equation~\eqref{logitunifiedmodel} in this example is $\boldsymbol{\eta}_x = {\mathbf X}_x \boldsymbol{\theta} = (\beta_1x, \ldots, \beta_{J-1}x, 0)^T$, where $\boldsymbol\theta = (\beta_1, \ldots, \beta_{J-1})^T \in \mathbb{R}^{J-1}$ are the model parameters. 
Then ${\mathbf U}_x = (u_{st}^x)_{s,t=1,\ldots, J}$ can be calculated from Theorem~A.2 in \cite{bu2020}. Note that $u_{sJ}^x = u_{Js}^x = 0$ for $s=1, \ldots, J-1$ and $u_{JJ}^x = 1$. The Fisher information matrix at ${\mathbf x}=x$ is ${\mathbf F}_x = {\mathbf X}_x^T {\mathbf U}_x {\mathbf X}_x = x^2 {\mathbf V}_x$, where ${\mathbf V}_x = (u_{st}^x)_{s,t=1,\ldots, J-1}$. Then $|{\mathbf F}_x| = x^{2(J-1)} |{\mathbf V}_x|$. According to Equation~(S.1) and Lemma~S.9 in the Supplementary Material of \cite{bu2020}, $|{\mathbf V}_x|$ equals to
\[
\left\{\begin{array}{ll}
\prod_{j=1}^J \pi_j^x & \mbox{ for baseline-category, }\\
& \mbox{  adjacent-categories,}\\
& \mbox{  and continuation-ratio}\\
\frac{\left[\prod_{j=1}^{J-1} \gamma_j^x (1-\gamma_j^x)\right]^2}{\prod_{j=1}^J \pi_j^x} & \mbox{ for cumulative logit models}
\end{array}\right.
\]
which is always positive, where $\gamma_j^x = \sum_{l=1}^j \pi_l^x \in (0,1)$, $j=1, \ldots, J-1$. In other words, ${\rm rank}({\mathbf F}_x)=p$ in this case, as long as $x\neq 0$.
}\hfill{$\Box$}
\end{example}
There also exists an example of a special MLM such that ${\mathbf F}_{{\mathbf x}} = {\mathbf F}_{{\mathbf x}'}$ but ${\mathbf x} \neq {\mathbf x}'$ (see Appendix~\ref{ex:F_x_1_x_2}).

To search for a new design point ${\mathbf x}^*$ in Step~$5^\circ$ of Algorithm~\ref{algo:ForLion_general}, we utilize the R function {\tt optim} with the option ``L-BFGS-B" that allows box constraints. L-BFGS-B is a limited-memory version of the BFGS algorithm, which itself is one of several quasi-Newton methods \citep{byrd1995}. It works fairly well in finding solutions even at the boundaries of the box constraints. We give explicit formulae for computing $d({\mathbf x}, \boldsymbol\xi)$ below and provide the first-order derivative of the sensitivity function for MLM~\eqref{logitunifiedmodel} in Appendix~\ref{sec:MLM_first_order}.

\begin{theorem}\label{thm:generalequivalence_mlm}
Consider MLM~\eqref{logitunifiedmodel} with a compact ${\mathcal X}$. A design $\boldsymbol\xi = \{({\mathbf x}_i, w_i), i=1, \ldots, m\}$ with $f({\boldsymbol\xi}) > 0$ is D-optimal if and only if $\max_{{\mathbf x} \in {\mathcal X}} d({\mathbf x}, {\boldsymbol\xi}) \leq p$, where 
\begin{eqnarray}
d({\mathbf x}, \boldsymbol\xi) &=& \sum_{j=1}^{J-1} u_{jj}^{\mathbf x} ({\mathbf h}_j^{\mathbf x})^T {\mathbf C}_{jj} {\mathbf h}_j^{\mathbf x} \nonumber\\
&+& \sum_{i=1}^{J-1} \sum_{j=1}^{J-1} u_{ij}^{\mathbf x}\cdot ({\mathbf h}_c^{\mathbf x})^T {\mathbf C}_{JJ} {\mathbf h}_c^{\mathbf x}\nonumber \\
&+& 2\sum_{i=1}^{J-2} \sum_{j=i+1}^{J-1} u_{ij}^{\mathbf x} ({\mathbf h}_j^{\mathbf x})^T {\mathbf C}_{ij} {\mathbf h}_i^{\mathbf x} \nonumber\\
&+& 2\sum_{i=1}^{J-1} \sum_{j=1}^{J-1} u_{ij}^{\mathbf x} ({\mathbf h}_c^{\mathbf x})^T {\mathbf C}_{iJ} {\mathbf h}_i^{\mathbf x}\label{eq:d(x,xi)_GLM}
\end{eqnarray}
${\mathbf C}_{ij} \in \mathbb{R}^{p_i\times p_j}$ is a submatrix of the $p\times p$ matrix
\footnotesize
\[
{\mathbf F}(\boldsymbol\xi)^{-1} =  \left[\begin{array}{ccc}
{\mathbf C}_{11} & \cdots & {\mathbf C}_{1J}\\
\vdots & \ddots & \vdots\\
{\mathbf C}_{J1} & \cdots & {\mathbf C}_{JJ}
\end{array}\right]
\]
\normalsize
$i,j=1, \ldots, J$, $p=\sum_{j=1}^J p_j$, and $p_J=p_c$~.
\hfill{$\Box$}
\end{theorem}

\subsection{Example: Emergence of house flies}\label{sec:MLM_example_fly}

In this section, we revisit the motivating example at the beginning of Section~\ref{sec:introduction}. The original design \citep{itepan1995} assigned $n_i=500$ pupae to each of $m=7$ doses, $x_i = 80, 100, 120, 140, 160, 180, 200$, respectively, which corresponds to the uniform design $\boldsymbol{\xi}_u$ in Table~\ref{tab:1}. Under a continuation-ratio non-proportional odds (npo) model adopted by \cite{atkinson1999}
\begin{eqnarray*}
\log\left(\frac{\pi_{i1}}{\pi_{i2} + \pi_{i3}}\right) &=& \beta_{11} + \beta_{12} x_i + \beta_{13} x_i^2\\ 
\log\left(\frac{\pi_{i2}}{\pi_{i3}}\right) &=& \beta_{21} + \beta_{22} x_i
\end{eqnarray*}
with fitted parameters $\boldsymbol\theta = (\beta_{11}, \beta_{12}, \beta_{13}, \beta_{21}, $ $\beta_{22})^T = (-1.935, -0.02642, 0.0003174,$ $-9.159,$ $0.06386)^T$, 
\cite{bu2020} obtained D-optimal designs under different grid sizes using their lift-one algorithm proposed for discrete factors. With grid size of $20$, that is, using the design space $\{80, 100, \ldots, 200\}$ that was evenly spaced by $20$ units, they obtained a design $\boldsymbol{\xi}_{20} $ containing four design points. With finer grid points on the same interval $[80, 200]$, both their grid-$5$ design $\boldsymbol{\xi}_5$ (with design space $\{80, 85, \ldots, 200\}$) and grid-$1$ design $\boldsymbol{\xi}_1$ (with design space $\{80, 81, \ldots, 200\}$) contain five design points (see Table~\ref{tab:1}). By incorporating the Fedorov-Wynn and lift-one algorithms and continuously searching the extended region $[0, 200]$, \cite{ai2023locally} obtained a four-points design $\boldsymbol\xi_{a}$ (see Example~S1 in their Supplementary Material).

For $x \in {\mathcal X} = [80, 200]$ as in \cite{bu2020}, our ForLion algorithm reports $\boldsymbol \xi_*$ with only three design points (see Table~\ref{tab:1}, as well as the Supplementary Material, Section~\ref{sec: MLM_fly_techdetail} for details). Compared with $\boldsymbol{\xi}_*$, the relative efficiencies of the $\boldsymbol{\xi}_u$, $\boldsymbol{\xi}_{20}$, $\boldsymbol{\xi}_5$, and $\boldsymbol{\xi}_1$, defined by $[f(\cdot)/f(\boldsymbol\xi_*)]^{1/5}$, are $82.79\%$, $99.68\%$, $99.91\%$, and $99.997\%$, respectively. The increasing pattern of relative efficiencies, from $\boldsymbol{\xi}_{20}$ to $\boldsymbol{\xi}_{5}$ and  $\boldsymbol{\xi}_{1}$, indicates that with finer grid points, the lift-one algorithm can search the design space more thoroughly and find better designs.
For $x \in [0,200]$ as in \cite{ai2023locally}, our algorithm yields $\boldsymbol \xi_*'$ with only three points (see Table~\ref{tab:1}) and the relative efficiency of \cite{ai2023locally}'s $\boldsymbol\xi_a$ with respect to $\boldsymbol\xi_*'$ is $99.81\%$.  Note that both $\boldsymbol{\xi}_*$ and $\boldsymbol{ \xi}_*'$ from our algorithm contain only three experimental settings, which achieves the minimum $m$ justified by \cite{bu2020}. Given that the cost of using the radiation device is expensive and each run of the experiment takes at least hours, our designs can save $40\%$ or $25\%$ cost and time compared with \cite{bu2020}'s and \cite{ai2023locally}'s, respectively. 

\begin{table}[hbt!]
\centering
\caption{Designs for the emergence of house files experiment}\label{tab:1}
{\tiny
			\begin{tabular}{|l|rrrrrrr|}
				\hline
				Design & $\boldsymbol{\xi}_u$ & $\boldsymbol{\xi}_{20}$ & $\boldsymbol{\xi}_5$ & $\boldsymbol{\xi}_1 $ & $\boldsymbol \xi_* $ & $\boldsymbol\xi_{a} $ & $\boldsymbol \xi_*'$ \\
				\hline 
                Range of $x$ & $[80, 200]$ & $[80, 200]$ & $[80, 200]$ & $[80, 200]$ & $[80, 200]$ & $[0, 200]$ & $[0, 200]$ \\ \hline
				&                            &                                 &                            &                              &                             & (0.00, 0.203)                  & (0.00, 0.203) \\
				& (80, 0.143)                & (80, 0.312)                    & (80, 0.316)               & (80, 0.316)                 &  (80.00, 0.316)            &                             &  \\
				&(100, 0.143)                &                                 &                            &                              &                             & (101.10, 0.397)              &  (103.56, 0.398) \\
				&(120, 0.143)                &  (120, 0.292)                  & (120, 0.143)              & (122, 0.079)                &  (122.78, 0.342)           &                             &         \\
				&(140, 0.143)                &  (140, 0.107)                  & (125, 0.200)              & (123, 0.264)                &                             & (147.80, 0.307)              &  (149.26, 0.399) \\
				&(160, 0.143)                &  (160, 0.290)                  & (155, 0.168)              & (157, 0.221)                &  (157.37, 0.342)           & (149.30, 0.093)              &         \\
				&(180, 0.143)                &                                 & (160, 0.172)              & (158, 0.121)                &                             &                          &         \\
				&(200, 0.143)                &                                 &                            &                              &                             &                          &         \\ \hline 
				Rela.Effi. & 82.79\%           &  99.68\%                          &  99.91\%                    &  99.997\%                     &   100\%                                  & 99.81\%                        &        100\%                    \\ 
				\hline
			\end{tabular}
}
\end{table}

\begin{figure*}[hbt!] 
    \centering
    \begin{minipage}[t]{0.45\textwidth}
        \centering
        \includegraphics[height=2.3in]{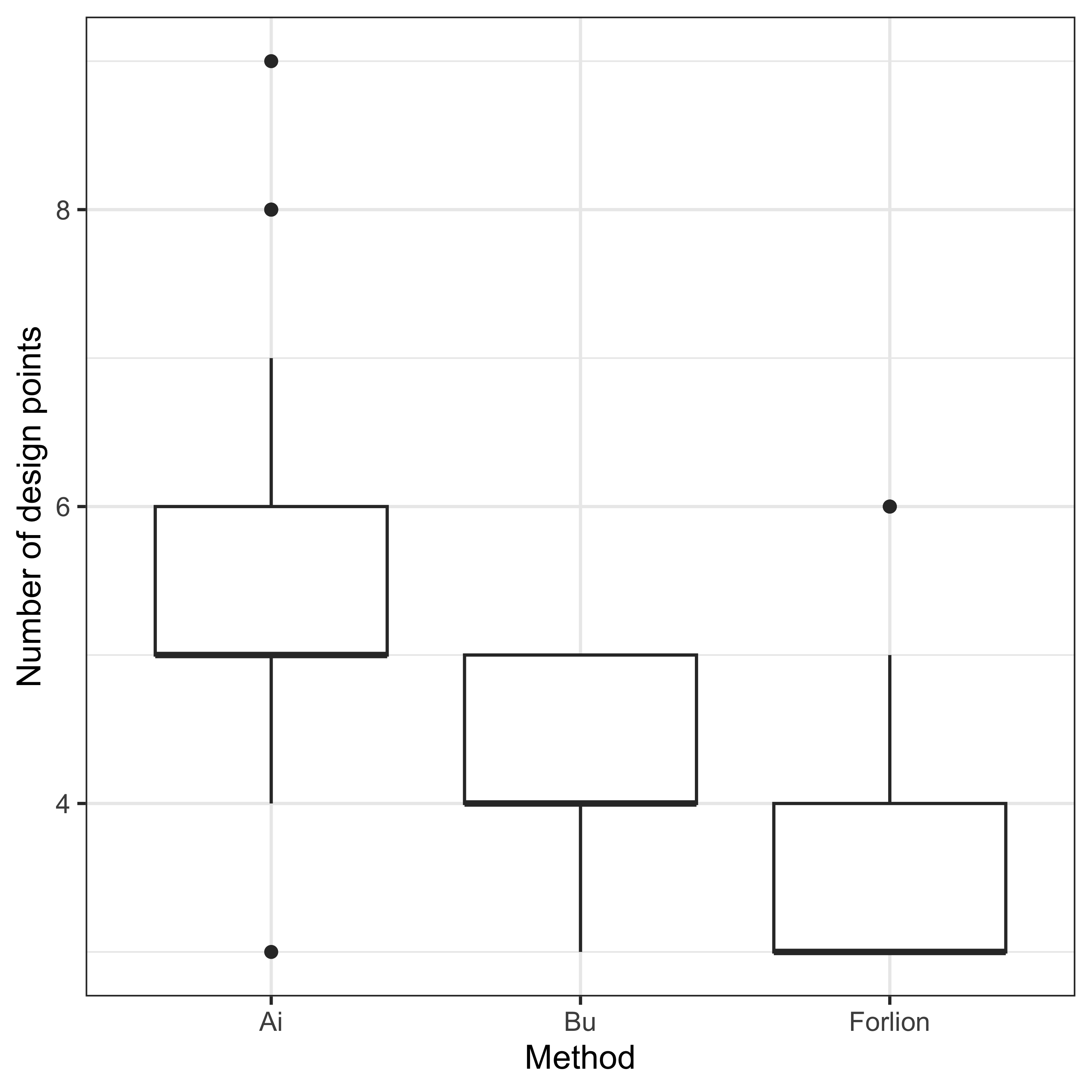}
        
       (a) Number of design points
    \end{minipage}%
    ~ 
    \begin{minipage}[t]{0.45\textwidth}
        \centering
        \includegraphics[height=2.3in]{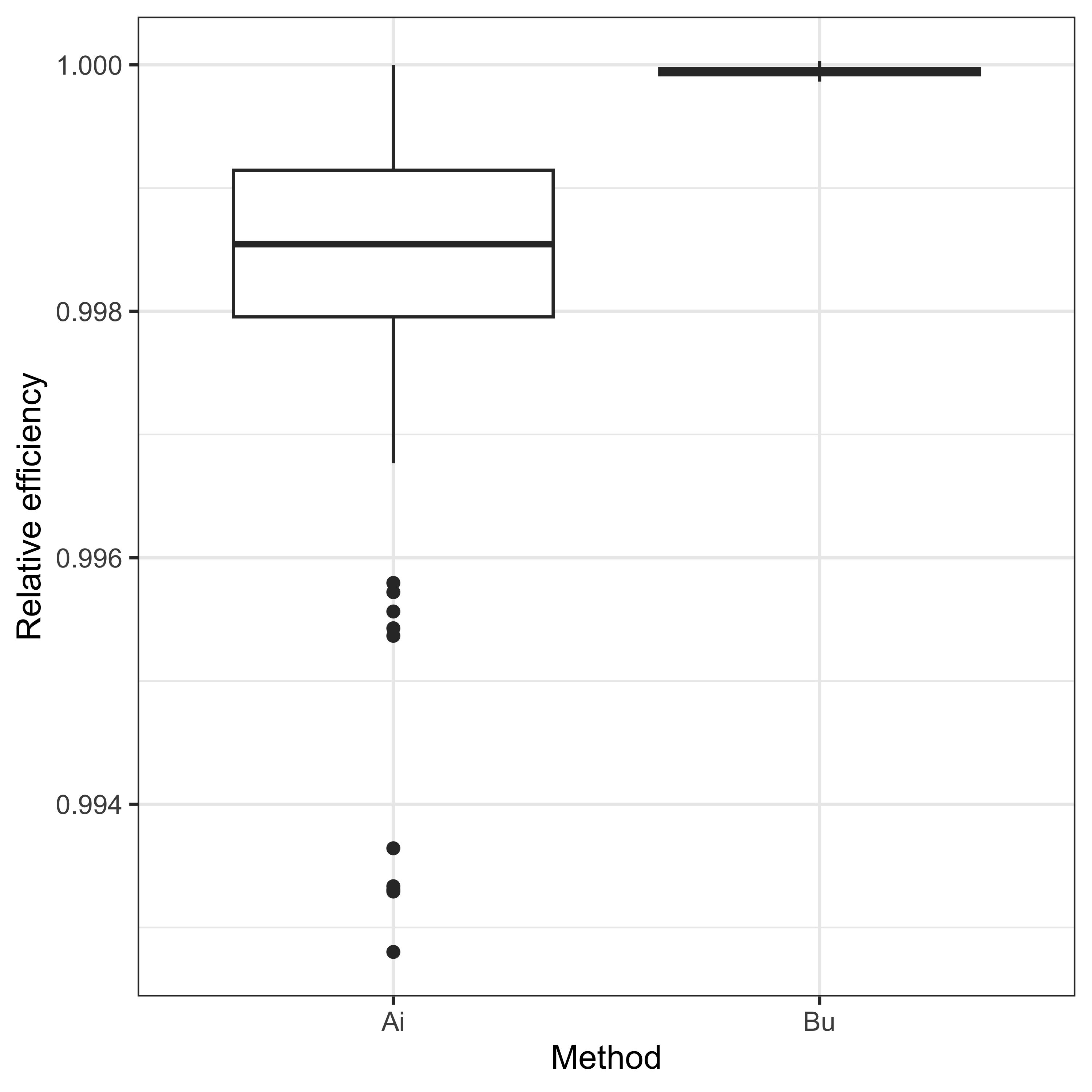}
        
       (b) Relative efficiency
    \end{minipage}
    \caption{\small Boxplots of 100 bootstrapped data for Ai's, Bu's, and ForLion's designs}\label{fig:mlm_fly}
\end{figure*}

To further check if the improvements by our designs are due to randomness, we conduct a simulation study by generating $100$ bootstrapped replicates from the original data. For each bootstrapped data, we obtain the fitted parameters, treat them as the true values, and obtain D-optimal designs by \cite{ai2023locally}'s, \cite{bu2020}'s grid-$1$, and our ForLion algorithm with $\delta = 0.1$ and $\epsilon=10^{-10}$, respectively.  As Figure~\ref{fig:mlm_fly} shows, our ForLion algorithm achieves the most efficient designs with the least number of design points. Actually, the median number of design points is $5$ for Ai's, $4$ for Bu's, and $3$ for ForLion's. Compared with ForLion's, the mean relative efficiency is $99.82\%$ for Ai's and $99.99\%$ for Bu's. As for computational time, the median time cost on a Windows 11 desktop with 32GB RAM and AMD Ryzen 7 5700G processor is $3.59$s for Ai's, $161.81$s for Bu's, and $44.88$s for ForLion's.

\section{D-optimal Designs for GLMs}\label{sec:GLM}

In this section, we consider experiments with a univariate response $Y$, which follows a distribution $f(y;\theta) = \exp\{y b(\theta) + c(\theta) + d(y)\}$ in the exponential family with a single parameter $\theta$. Examples include binary response $Y\sim$ Bernoulli$(\theta)$, count response $Y\sim$ Poisson$(\theta)$, positive response $Y\sim$ Gamma$(\kappa, \theta)$ with known $\kappa>0$, and continuous response $Y\sim N(\theta, \sigma^2)$ with known $\sigma^2 > 0$ \citep{pmcc1989}.
Suppose independent responses $Y_1, \ldots, Y_n$ are collected with corresponding factor level combinations ${\mathbf x}_1, \ldots, {\mathbf x}_n \in {\mathcal X} \subset \mathbb{R}^{d}$, where ${\mathbf x}_i = (x_{i1}, \ldots, x_{id})^T$. Under a generalized linear model (GLM), there exist a link function $g$, parameters of interest $\boldsymbol{\beta} = (\beta_1, \beta_2, \ldots, \beta_p)^T$, and the corresponding vector of $p$ known and deterministic predictor functions ${\mathbf h} = (h_1, \ldots, h_p)^T$, such that
\begin{equation}\label{eq:glm}
E(Y_i) = \mu_i\mbox{ and } \eta_i = g(\mu_i)= {\mathbf X}_i^T\boldsymbol{\beta}
\end{equation}
where ${\mathbf X}_i = {\mathbf h}({\mathbf x}_i) = (h_1({\mathbf x}_i), \ldots, h_p({\mathbf x}_i))^T$, $i=1, \ldots, n$. For many applications, $h_1({\mathbf x}_i)\equiv 1$ represents the intercept of the model.

\subsection{ForLion algorithm specialized for GLM}\label{sec:GLM:forlion}

Due to the specific form of GLM's Fisher information (see Section~\ref{sec:GLM_Fisher} and \eqref{eq:Fisher_GLM} in the Supplementary Material), the lift-one algorithm can be extremely efficient by utilizing analytic solutions for each iteration \citep{ym2015}. In this section, we specialize the ForLion algorithm for GLM with explicit formulae in Steps~$3^\circ$, $5^\circ$, and $6^\circ$.

For GLM~\eqref{eq:glm}, our goal is to find a design $\boldsymbol\xi = \{({\mathbf x}_i, w_i), i=1, \ldots, m\}$ maximizing $f({\boldsymbol\xi}) = |{\mathbf X}_{\boldsymbol\xi}^T{\mathbf W}_{\boldsymbol\xi}{\mathbf X}_{\boldsymbol\xi}|$, where ${\mathbf X}_{\boldsymbol\xi} = ({\mathbf h}({\mathbf x}_1), \ldots, {\mathbf h}({\mathbf x}_m))^T \in \mathbb{R}^{m\times p}$ with known predictor functions $h_1, \ldots, h_p$~, and ${\mathbf W}_{\boldsymbol\xi} = {\rm diag}\{w_1 \nu({\boldsymbol\beta}^T {\mathbf h}({\mathbf x}_1)), \ldots, w_m \nu({\boldsymbol\beta}^T {\mathbf h}({\mathbf x}_m))\}$ with known parameters $\boldsymbol\beta = (\beta_1, \ldots, \beta_p)^T$ and a positive differentiable function $\nu$, where $\nu(\eta_i) = (\partial \mu_i/\partial\eta_i)^2/{\rm Var}(Y_{i})$, for $i=1, \ldots, m$ (see Sections~\ref{subsec:glm.formula} and \ref{sec:GLM_Fisher} in the Supplementary Material for examples and more technical details). The sensitivity function $d({\mathbf x}, {\boldsymbol\xi}) = {\rm tr}({\mathbf F}({\boldsymbol\xi})^{-1} {\mathbf F}_{\mathbf x})$ in Step~$5^\circ$ of Algorithm~\ref{algo:ForLion_general} can be written as $\nu({\boldsymbol\beta}^T {\mathbf h}({\mathbf x})) \cdot {\mathbf h}({\mathbf x})^T ({\mathbf X}_{\boldsymbol\xi}^T{\mathbf W}_{\boldsymbol\xi}{\mathbf X}_{\boldsymbol\xi})^{-1}$ $ {\mathbf h}({\mathbf x})$. 
As a direct conclusion of the general equivalence theorem (see Theorem~2.2 in \cite{fedorov2014}), we have the following theorem for GLMs.

\begin{theorem}\label{thm:generalequivalence}
Consider GLM~\eqref{eq:glm} with a compact design region ${\mathcal X}$. A design $\boldsymbol\xi = \{({\mathbf x}_i, w_i), i=1, \ldots, m\}$ with $f({\boldsymbol\xi}) = |{\mathbf X}_{\boldsymbol\xi}^T{\mathbf W}_{\boldsymbol\xi}{\mathbf X}_{\boldsymbol\xi}| > 0$ is D-optimal if and only if
\begin{equation}\label{eq:equivalence}
\max_{{\mathbf x} \in {\mathcal X}} \nu({\boldsymbol\beta}^T {\mathbf h}({\mathbf x})) \cdot {\mathbf h}({\mathbf x})^T ({\mathbf X}_{\boldsymbol\xi}^T{\mathbf W}_{\boldsymbol\xi}{\mathbf X}_{\boldsymbol\xi})^{-1} {\mathbf h}({\mathbf x})  \leq p
\end{equation}
\end{theorem}

Given the design ${\boldsymbol\xi}_t = \{({\mathbf x}_i^{(t)}, w_i^{(t)}), i=1, \ldots, m_t\}$ at the $t$th iteration, suppose in Step~$5^\circ$ we find the design point ${\mathbf x}^*  \in {\mathcal X}$ maximizing 
\[
d({\mathbf x}, {\boldsymbol\xi}_t) = \nu({\boldsymbol\beta}^T {\mathbf h}({\mathbf x})) \cdot {\mathbf h}({\mathbf x})^T ({\mathbf X}_{{\boldsymbol\xi}_t}^T{\mathbf W}_{{\boldsymbol\xi}_t}{\mathbf X}_{{\boldsymbol\xi}_t})^{-1} {\mathbf h}({\mathbf x})
\]
according to Theorem~\ref{thm:generalequivalence}. Recall that $d({\mathbf x}^*, {\boldsymbol\xi}_t) \leq p$ in this step implies the optimality of ${\boldsymbol\xi}_t$ and the end of the iterations. If $d({\mathbf x}^*, {\boldsymbol\xi}_t) > p$, ${\mathbf x}^*$ will be added to form the updated design ${\boldsymbol{\xi}}_{t+1}$~. For GLMs, instead of letting ${\boldsymbol\xi}_{t+1} = {\boldsymbol\xi}_t \bigcup \{({\mathbf x}^*, 0)\}$, we recommend $\{({\mathbf x}_i^{(t)}, (1-\alpha_t)w_i^{(t)}), i=1, \ldots, m_t\} \bigcup \{({\mathbf x}^*, \alpha_t)\}$, denoted by $(1-\alpha_t) {\boldsymbol\xi}_t \bigcup \{({\mathbf x}^*, \alpha_t)\}$ for simplicity, where $\alpha_t \in [0,1]$ is an initial allocation for the new design point $\mathbf x^*$, which is determined by Theorem~\ref{lemma:alphat}.

\begin{theorem}\label{lemma:alphat}
Given ${\boldsymbol\xi}_t = \{({\mathbf x}_i^{(t)}, w_i^{(t)}), i=1, \ldots, m_t\}$ and ${\mathbf x}^* \in {\mathcal X}$, if we consider ${\boldsymbol\xi}_{t+1}$ in the form of $(1-\alpha) {\boldsymbol\xi}_t \bigcup \{({\mathbf x}^*, \alpha)\}$ with $\alpha \in [0,1]$, then
\[
\alpha_t = \left\{
\begin{array}{cl}
\frac{2^p\cdot d_t - (p+1) b_t}{p(2^p\cdot d_t - 2 b_t)} & \mbox{if }2^p\cdot d_t > (p+1) b_t\\
0 & \mbox{otherwise}
\end{array}
\right.
\]
maximizes $f({\boldsymbol\xi}_{t+1})$ with $d_t = f(\{({\mathbf x}^{(t)}_1, w^{(t)}_1/2),$ $\ldots, ({\mathbf x}^{(t)}_{m_t}, w^{(t)}_{m_t}/2), ({\mathbf x}^*, 1/2)\})$ and $b_t = f(\boldsymbol\xi_t)$.
\end{theorem}

Based on $\alpha_t$ in Theorem~\ref{lemma:alphat}, which is obtained essentially via one iteration of the lift-one algorithm \citep{ymm2016,ym2015} with ${\mathbf x}^*$ added, we update Step~$6^\circ$ of Algorithm~\ref{algo:ForLion_general} with Step~$6'$ for GLMs, which speeds up the ForLion algorithm significantly. 
\begin{itemize}
    \item[$6'$] If $d({\mathbf x}^*, {\boldsymbol\xi}_t) \leq p$, go to Step~$7^\circ$. Otherwise, we let ${\boldsymbol\xi}_{t+1} = (1-\alpha_t){\boldsymbol\xi}_t \bigcup \{({\mathbf x}^*, \alpha_t)\}$, replace $t$ by $t+1$, and go back to Step~$2^\circ$, where $\alpha_t$ is given by Theorem~\ref{lemma:alphat}.
\end{itemize}

The advantages of the lift-one algorithm over commonly used numerical algorithms include simplified computation and exact zero weight for negligible design points. For GLMs, Step~$3^\circ$ of Algorithm~\ref{algo:ForLion_general} should be specialized with analytic iterations as in \cite{ym2015}. We provide the explicit formula for the sensitivity function's first-order derivative in Section~\ref{sec:GLM_sensitivity_derivative} of the Supplementary Material for Step~$5^\circ$.

By utilizing the analytical solutions for GLM in Steps~$3^\circ$, $5^\circ$ and $6'$, the computation is much faster than the general procedure of the ForLion algorithm (see Example~\ref{ex:stufken} in Section~\ref{sec:maineffects}).

\subsection{Minimally supported design and initial design}\label{sec:minimally_supported}

A minimally supported design $\boldsymbol\xi = \{({\mathbf x}_i, w_i), i=1, \ldots, m\}$ achieves the smallest possible $m$ such that $f({\boldsymbol\xi}) > 0$, or equivalently, the Fisher information matrix is of full rank. Due to the existence of Example~\ref{ex:rank_F_p}, $m$ could be as small as $1$ for an MLM. Nevertheless, $m$ must be $p$ or above for GLMs due to the following theorem.

\begin{theorem}\label{thm:minimallysupported}
Consider a design $\boldsymbol\xi = \{({\mathbf x}_i, w_i), i=1, \ldots, m\}$ with $m$ support points, that is, $w_i > 0$, $i=1, \ldots, m$, for GLM~\eqref{eq:glm}. Then $f({\boldsymbol\xi}) > 0$ only if ${\mathbf X}_{\boldsymbol\xi}$ is of full column rank $p$, that is, ${\rm rank}({\mathbf X}_{\boldsymbol\xi}) = p$. Therefore, a minimally supported design contains at least $p$ support points. Furthermore, if $\nu({\boldsymbol\beta}^T {\mathbf h}({\mathbf x})) > 0$ for all ${\mathbf x}$ in the design region ${\mathcal X}$, then $f({\boldsymbol\xi}) > 0$ if and only if ${\rm rank}({\mathbf X}_{\boldsymbol\xi}) = p$.
\end{theorem}

Theorem~\ref{thm:minimalluniform} shows that a minimally supported D-optimal design under a GLM must be a uniform design.

\begin{theorem}\label{thm:minimalluniform}
Consider a minimally supported design $\boldsymbol\xi = \{({\mathbf x}_i, w_i), i=1, \ldots, p\}$ for GLM~\eqref{eq:glm} that satisfies $f({\boldsymbol\xi}) > 0$. It is D-optimal only if $w_i=p^{-1}$, $i=1, \ldots, p$. That is, it is a uniform allocation on its support points.
\end{theorem}

Based on Theorem~\ref{thm:minimalluniform}, we recommend a minimally supported design as the initial design for the ForLion algorithm under GLMs. The advantage is that once $p$ design points ${\mathbf x}_1, \ldots, {\mathbf x}_p$ are chosen from ${\mathcal X}$, such that the model matrix ${\mathbf X}_{\boldsymbol\xi} = ({\mathbf h}({\mathbf x}_1), \ldots, {\mathbf h}({\mathbf x}_p))^T$ is of full rank $p$, then the design $\boldsymbol\xi = \{({\mathbf x}_i, 1/p), i=1, \ldots, p\}$ is D-optimal given those $p$ design points.

Recall that a typical design space can take the form of ${\mathcal X} = \prod_{j=1}^d I_j$, where $I_j$ is either a finite set of distinct numerical levels or an interval $[a_j, b_j]$, and $a_j = \min I_j$ and $b_j = \max I_j$ even if $I_j$ is a finite set. As one option in Step~$1^\circ$ of Algorithm~\ref{algo:ForLion_general}, we suggest to choose $p$ initial design points from $\prod_{j=1}^d \{a_j, b_j\}$.
For typical applications, we may assume that there exist $p$ distinct points in $\prod_{j=1}^d \{a_j, b_j\}$ such that the corresponding model matrix ${\mathbf X}_{\boldsymbol\xi}$ is of full rank, or equivalently, the $2^d\times p$ matrix consisting of rows ${\mathbf h}({\mathbf x})^T, {\mathbf x} \in \prod_{j=1}^d \{a_j, b_j\}$ is of full rank $p$. Herein, we specialize Step~$1^\circ$ of Algorithm~\ref{algo:ForLion_general} for GLMs as follows:

\begin{itemize}
\item[$1'$] Construct an initial design ${\boldsymbol\xi}_0 = \{({\mathbf x}_i^{(0)}, p^{-1}),$ $i=1, \ldots, p\}$ such that ${\mathbf x}_1^{(0)},$ $\ldots,$ ${\mathbf x}_p^{(0)} \in \prod_{j=1}^d \{a_j, b_j\}$ and ${\mathbf X}_{\boldsymbol{\xi}_0} = ({\mathbf h}({\mathbf x}_1^{(0)}), \ldots, {\mathbf h}({\mathbf x}_p^{(0)}))^T$ is of full rank $p$.
\end{itemize}

\subsection{Examples under GLMs}\label{sec:maineffects}

In this section, we use two examples to show the performance of our ForLion algorithm under GLMs, both with continuous factors involved only as main effects, which have simplified notations (see Section~\ref{sec:maineffects_supp} in the Supplementary Material).

\begin{example}\label{ex:stufken}{\rm
In Example~4.7 of \cite{stufken2012}, they considered a logistic model with three continuous factors ${\rm logit}(\mu_i) = \beta_0 + \beta_1 x_{i1} + \beta_2 x_{i2} + \beta_3 x_{i3}$ with $x_{i1} \in [-2, 2]$, $x_{i2} \in [-1,1]$, and $x_{i3} \in (-\infty, \infty)$.	Assuming $(\beta_0, \beta_1, \beta_2, \beta_3) = (1, -0.5, 0.5, 1)$, they obtained an $8$-points D-optimal design ${\boldsymbol\xi}_o$ theoretically. Using a MacOS-based laptop with CPU 2 GHz Quad-Core and memory 16GB 3733 MHz, our ForLion algorithm specialized for GLMs takes 17 seconds and our general ForLion algorithm takes 124 seconds to get the same design $\boldsymbol\xi_*$ (see Table~\ref{tab:3continuous}).
The relative efficiency of $\boldsymbol\xi_*$ compared with $\boldsymbol\xi_o$ is simply $100\%$.
Note that $\boldsymbol\xi_o$ was obtained from an analytic approach requiring an unbounded $x_{i3}$~. For bounded $x_{i3}$~, such as $x_{i3}\in [-1, 1]$, $[-2, 2]$, or $[-3, 3]$, we can still use the ForLion algorithm to obtain the corresponding D-optimal designs, whose relative efficiencies compared with $\boldsymbol\xi_o$ or $\boldsymbol\xi_*$ are $85.55\%$, $99.13\%$, and $99.99993\%$, respectively.
\hfill{$\Box$}
}
\end{example}

\begin{figure*}[htb!] 
    \centering
    \begin{minipage}[t]{0.45\textwidth}
        \centering
        \includegraphics[height=2.3in]{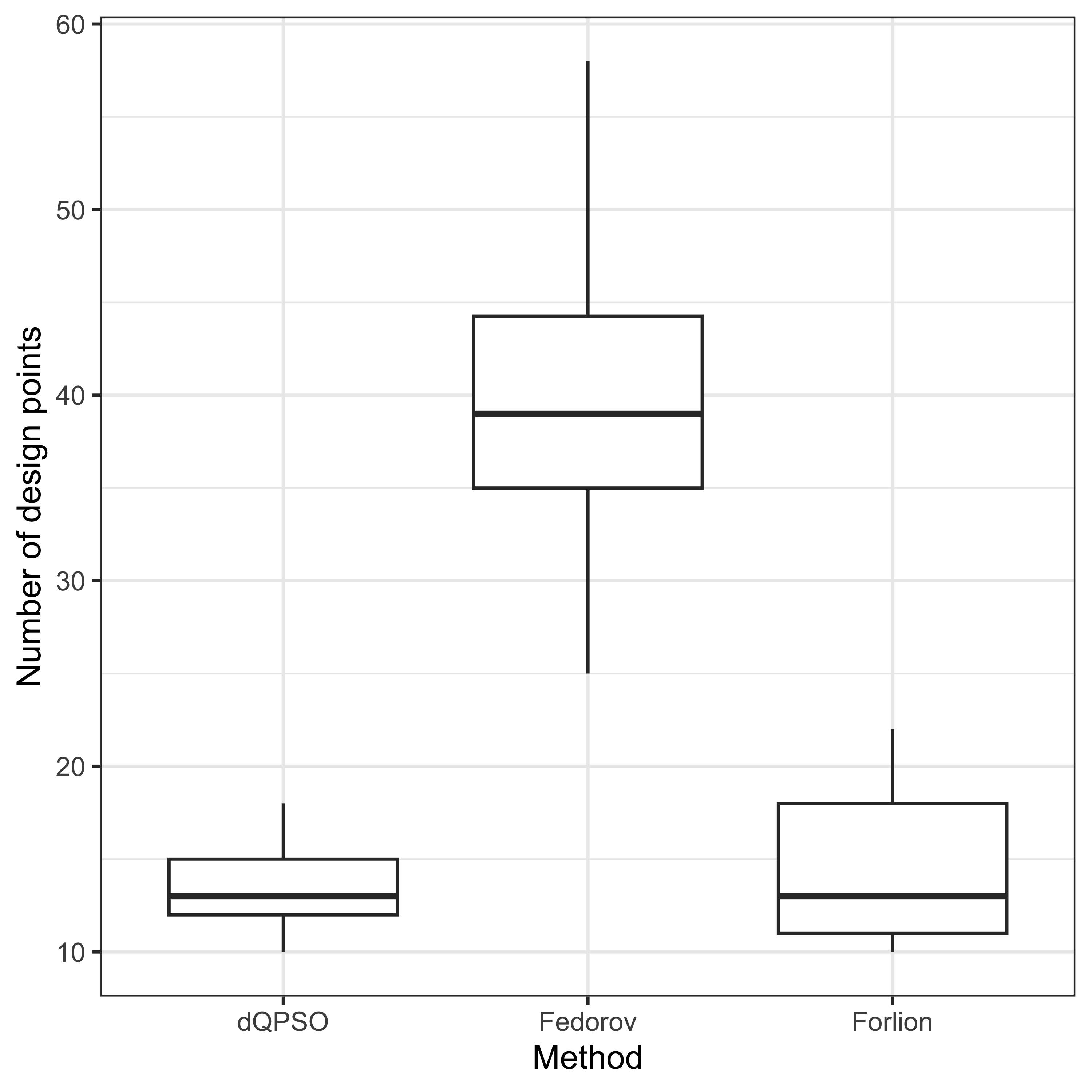}
        
        (a) Number of design points
    \end{minipage}%
    ~ 
    \begin{minipage}[t]{0.45\textwidth}
        \centering
        \includegraphics[height=2.3in]{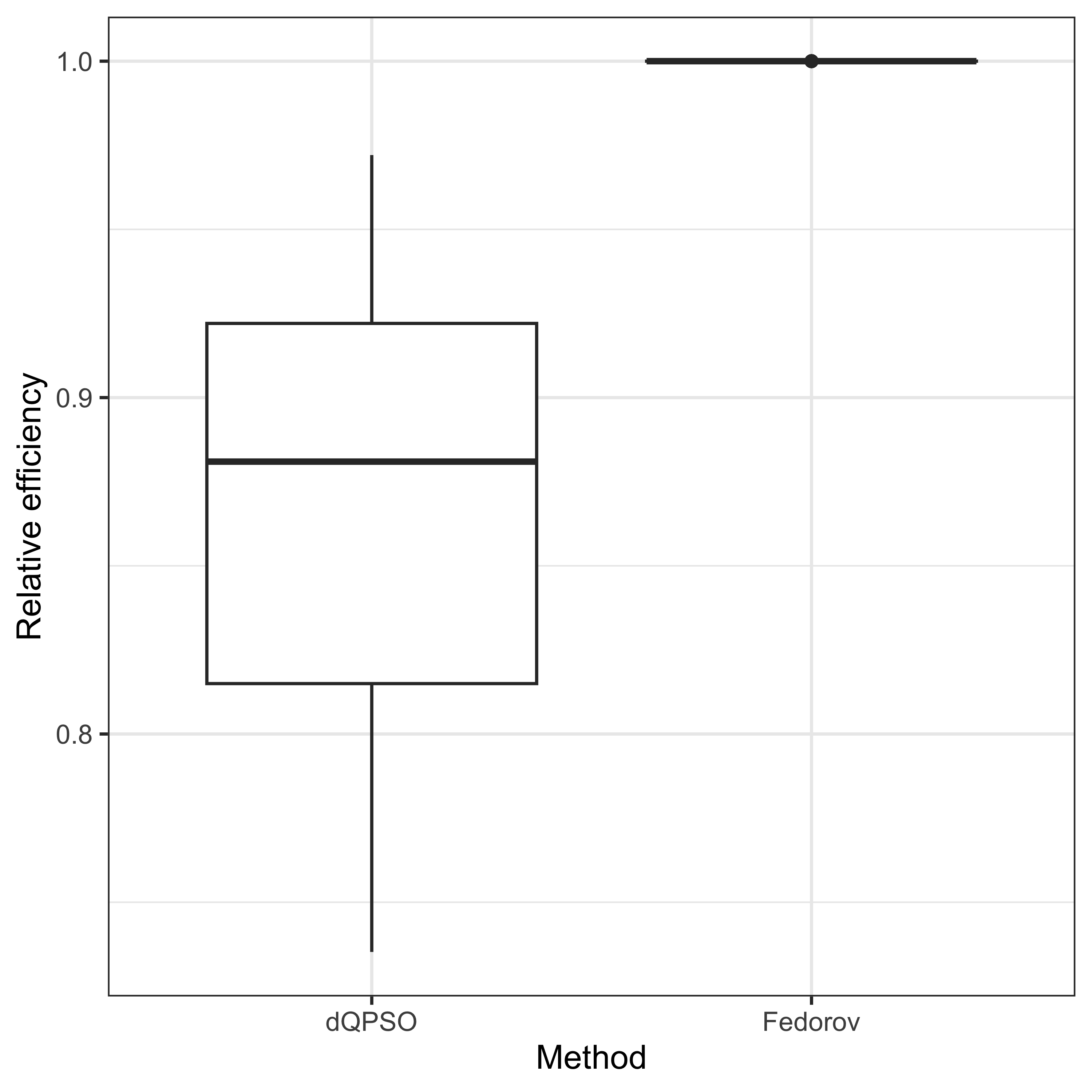}
     
       (b) Relative efficiency w.r.t. ForLion
    \end{minipage}
    \caption{\small Boxplots of 100 simulations for $d$-QPSO, Fedorov-Wynn-liftone, and ForLion algorithms for the electrostatic discharge experiment (Example~\ref{ex:ESD})}\label{fig:glm_ESD}
\end{figure*}

\begin{example}\label{ex:ESD}{\rm
\cite{lukemire2018} reconsidered the electrostatic discharge (ESD) experiment described by \cite{whitman2006} with a binary response and five mixed factors. The first four factors {\tt LotA}, {\tt LotB}, {\tt ESD}, {\tt Pulse} take values in $\{-1, 1\}$, and the fifth factor {\tt Voltage} $\in [25, 45]$ is continuous. Using their $d$-QPSO algorithm, \cite{lukemire2018} obtained a $13$-points design $\boldsymbol\xi_o$ for the model ${\rm logit} (\mu) = \beta_0 + \beta_1 {\tt Lot A} + \beta_2 {\tt Lot B} + \beta_3 {\tt ESD} + \beta_4 {\tt Pulse} + \beta_5 {\tt Voltage} + \beta_{34} ({\tt ESD}\times {\tt Pulse})$ with assumed parameter values $\boldsymbol\beta = (-7.5, 1.50, -0.2, -0.15,$ $0.25, 0.35, 0.4)^T$. It takes $88$ seconds using the same laptop as in Example~\ref{ex:stufken} for our (GLM) ForLion algorithm to find a slightly better design $\boldsymbol\xi_*$ consisting of $14$ points (see Table~\ref{tab:ESD} in the Supplementary Material) with relative efficiency $[f(\boldsymbol\xi_*)/f(\boldsymbol\xi_o)]^{1/7} = 100.08\%$. 

\begin{table*}[hbt!]
\centering
\caption{Designs obtained for Example~\ref{ex:stufken}}\label{tab:3continuous}
{\footnotesize
 \begin{tabular}{|c|rrrr|rrrr| }\hline
		& \multicolumn{4}{c|}{\citeauthor{stufken2012}'s ${\boldsymbol\xi}_o$} & \multicolumn{4}{c|}{ForLion's ${\boldsymbol\xi}_*$}\\ \hline
	$i$ & $x_{i1}$ & $x_{i2}$ & $x_{i3}$ & $w_i$ & $x_{i1}$ & $x_{i2}$ & $x_{i3}$ & $w_i$\\ \hline
1  & -2  & -1  & -2.5436  & 0.125  & -2  & -1  & -2.5433  & 0.125 \\
2  & -2  & -1  & -0.4564  & 0.125  & -2  & -1  & -0.4565  & 0.125 \\
3  & -2  & 1  & -3.5436  & 0.125  & -2  & 1  & -3.5433  & 0.125 \\
4  & -2  & 1  & -1.4564  & 0.125  & -2  & 1  & -1.4564  & 0.125 \\
5  & 2  & -1  & -0.5436  & 0.125  & 2  & -1  & -0.5435  & 0.125 \\
6  & 2  & -1  & 1.5436  & 0.125  & 2  & -1  & 1.5434  & 0.125 \\
7  & 2  & 1  & -1.5436  & 0.125  & 2  & 1  & -1.5445  & 0.125 \\
8  & 2  & 1  & 0.5436  & 0.125  & 2  & 1  & 0.5433  & 0.125 \\
		\hline
    \end{tabular}	
}\end{table*}

To make a thorough comparison, we randomly generate 100 sets of parameters $\boldsymbol{\beta}$ from independent uniform distributions: $U(1.0, 2.0)$ for {\tt LotA}, $U(-0.3,$ $ -0.1)$ for {\tt LotB}, $U(-0.3, 0.0)$ for {\tt ESD}, $U(0.1, 0.4)$ for {\tt Pulse}, $U(0.25, 0.45)$ for {\tt Voltage}, $U(0.35, 0.45)$ for {\tt ESD}$\times${\tt Pulse}, and $U(-8.0, -7.0)$ for Intercept. For each simulated $\boldsymbol{\beta}$, we treat it as the true parameter values and obtain D-optimal designs using $d$-QPSO, ForLion, and Fedorov-Wynn-liftone (that is, the ForLion algorithm without the merging step, similarly in spirit to \cite{ai2023locally}'s) algorithms. For the ForLion algorithm, we use $\delta=0.03$ and $\epsilon=10^{-8}$ (see Section~\ref{sec:mergedistance_delta} in the Supplementary Material for more discussion on choosing $\delta$). For the $d$-QPSO algorithm, following \cite{lukemire2018}, we use $5$ swarms with $30$ particles each and the algorithm searches design with up to $18$ support points and the maximum number of iterations $4,000$. 
Figure~\ref{fig:glm_ESD} shows the numbers of design points of the three algorithms and the relative efficiencies. The median number of design points is $13$ for $d$-QPSO's, $39$ for Fedorov-Wynn-liftone's, and $13$ for Forlion's. The mean relative efficiencies compared with our ForLion D-optimal designs, defined as $[f(\cdot)/f(\boldsymbol\xi_{\rm ForLion})]^{1/7}$, is $86.95\%$ for $d$-QPSO's and $100\%$ for Fedorov-Wynn-liftone's. The median running time on the same desktop in Section~\ref{sec:MLM_example_fly} is $10.94$s for $d$-QPSO's, $129.74$s for Fedorov-Wynn-liftone's, and $71.21$s for ForLion's.
	
 \hfill{$\Box$}
	} 
\end{example}

\section{Conclusion and Discussion}\label{sec:conclusion}

In this paper, we develop the ForLion algorithm to find locally D-optimal approximate designs under fairly general parametric models with mixed factors. 

Compared with \cite{bu2020}'s and \cite{ai2023locally}'s algorithm, our ForLion algorithm can reduce the number of distinct experimental settings by $25\%$ above on average while keeping the highest possible efficiency (see Section~\ref{sec:MLM_example_fly}). In general, for experiments such as the Emergence of house flies (see Section~\ref{sec:MLM_example_fly}), the total experimental cost not only relies on the number of experimental units, but also the number of distinct experimental settings (or runs). For such kind of experiments, an experimental plan with fewer distinct experimental settings may allow the experimenter to support more experimental units under the same budget limit, and complete the experiment with less time cost. Under the circumstances, our ForLion algorithm may be extended for a modified design problem that maximizes $n^p|{\mathbf F}(\boldsymbol{\xi})|$ under a budget constraint $mC_r + nC_u \leq C_0$, that incorporates the experimental run cost $C_r$ and the experimental unit cost $C_u$, and perhaps a time constraint $m\leq M_0$ as well.

Compared with PSO-type algorithms, the ForLion algorithm can improve the relative efficiency by $17.5\%$ above on average while achieving a low number of distinct experimental settings (see Example~\ref{ex:ESD}). Our ForLion algorithm may be extended for other optimality criteria by adjusting the corresponding objective function $f(\boldsymbol{\xi})$ and the sensitivity function $d(\mathbf{x}, \boldsymbol{\xi})$, as well as a lift-one algorithm modified accordingly to align with those criteria. 

In Step~$5^\circ$ of Algorithm~\ref{algo:ForLion_general}, the search for a new design point $x^*$ involves solving for continuous design variables for each level combination of the discrete variables. When the number of discrete variables increases, the number of scenarios grows exponentially, which may cause computationally inefficiency. In this case, one possible solution is to treat some of the discrete variables as continuous variables first, run the ForLion algorithm to obtain a design with continuous levels of those discrete factors, and then modify the design points by rounding the continuous levels of discrete factors to their nearest feasible levels. For a binary factor, one may simply use $[0,1]$ or $[-1,1]$ as a continuous region of the possible levels. Note that there are also experiments with discrete factors involving three or more categorical levels. For example, the factor of Cleaning Method in an experiment on a polysilicon deposition process for manufacturing very large scale integrated circuits has three levels, namely, None, CM$_2$, and CM$_3$ \citep{phadke1989}. For those scenarios, one may first transform such a discrete factor, say, the Cleaning Method, to two dummy variables (that is, the indicator variables for CM$_2$ and CM$_3$, respectively) taking values in $\{0,1\}$, then run the ForLion algorithm by treating them as two continuous variables taking values in $[0,1]$.

\section*{Supplementary Information}

We provide explicit formulae of $u_{st}^{\mathbf x}$ in the Fisher information $\mathbf F_{\mathbf x}$ for MLM~\eqref{logitunifiedmodel} in Appendix~\ref{sec:computing_F_x}, a special example of MLM that ${\mathbf F}_{{\mathbf x}} = {\mathbf F}_{{\mathbf x}'}$ in Appendix~\ref{ex:F_x_1_x_2}, and the first-order derivative of $d(\mathbf x, \boldsymbol \xi)$ for MLM~\eqref{logitunifiedmodel} in Appendix~\ref{sec:MLM_first_order}.

\subsection*{Supplementary Material}

Contents of the Supplementary Material are listed below:
{\bf S.1 Commonly used GLMs:} A list of commonly used GLM models, corresponding link functions, $\nu$ functions, and their first-order derivatives;
{\bf S.2 Technical details of house flies example:} Technical details of applying the ForLion algorithm to the emergence of house flies example;
{\bf S.3 Example: Minimizing surface defects:} An example with cumulative logit po model that shows the advantages of the ForLion algorithm;
{\bf S.4 Fisher information matrix for GLMs:} Formulae for computing Fisher information matrix for GLMs;
{\bf S.5 First-order derivative of sensitivity function for GLMs:} Formulae of $\partial d({\mathbf x}, \boldsymbol\xi)/\partial x_i$ for GLMs;
{\bf S.6 GLMs with main-effects continuous factors:} Details of GLMs with main-effects continuous factors;
{\bf S.7 Electrostatic discharge example supplementary:} The optimal design table for electrostatic discharge example and a simulation study on the effects of merging threshold $\delta$;
{\bf S.8 Assumptions needed for Theorem~\ref{thm:doptimality}};
{\bf S.9 Proofs:} Proofs for theorems in this paper.



\medskip\noindent

\subsection*{Acknowledgements}

The authors gratefully acknowledge the authors of \cite{ai2023locally}, \cite{lukemire2018} and \cite{lukemire2022} for kindly sharing their source codes, which we used to implement and compare their methods with ours. The authors gratefully acknowledge the support from the U.S.~NSF grants DMS-1924859 and DMS-2311186.

\section*{Author Information}

\subsection*{Authors and Affiliations}
{\bf Department of Mathematics, Statistics, and Computer Science, University of Illinois at Chicago, Chicago, IL, 60607, USA\\}
Yifei Huang $\&$ Jie Yang\\
\hfil\par
\noindent{\bf Department of Mathematics, University of Alabama at Birmingham, Birmingham, AL, 35294, USA\\}
Keren Li\\
\hfil\par
\noindent{\bf Department of Statistics, University of Georgia, Athens, GA, 30602, USA\\}
Abhyuday Mandal

\subsection*{Contributions}

Conceptualization and methodology, all authors; software, Y.H., K.L., and J.Y.; validation, Y.H., A.M., and J.Y.; formal analysis, Y.H. and K.L.; investigation, Y.H. and J.Y.; resources, all authors; data curation, Y.H.; writing--original draft preparation, all authors; writing--review and editing, all authors.; supervision, J.Y. and A.M.; project administration, J.Y. and A.M.; funding acquisition, J.Y. and A.M.. All authors have read and agreed to the published version of the manuscript.

\subsection*{Corresponding author}
Correspondence to \href{mailto:jyang06@uic.edu}{Jie Yang}

\begin{appendices}

\vspace{1cm}
\noindent
{\Large\bf Appendices}

\renewcommand{\theequation}{\thesection\arabic{equation}}
\setcounter{equation}{0}

\section{Computing $u_{st}^{\mathbf x}$ in Fisher information $\mathbf F_{\mathbf x}$}\label{sec:computing_F_x}

In this section, we provide more technical details for Section~\ref{sec:F_x_for_MLM} and Theorem~\ref{thm:F_x}.

For MLM \eqref{logitunifiedmodel}, Corollary~3.1 in \cite{bu2020} provided an alternative form ${\mathbf F}_{{\mathbf x}_i} = {\mathbf X}_i^T {\mathbf U}_i {\mathbf X}_i$, which we use for computing the Fisher information ${\mathbf F}_{\mathbf x}$ at an arbitrary ${\mathbf x} \in {\mathcal X}$.
More specifically, 
first of all,
the corresponding model matrix at ${\mathbf x}$ is
\footnotesize
\begin{equation}\label{eq:X_x}
{\mathbf X}_{\mathbf x}= \begin{pmatrix}
 {\mathbf h}_1^T({\mathbf x}) &  \boldsymbol0^T & \cdots & \boldsymbol0^T& {\mathbf h}_c^T({\mathbf x})\\
 \boldsymbol0^T &  {\mathbf h}_2^T({\mathbf x}) &\ddots & \vdots & \vdots\\
\vdots &  \ddots& \ddots &  \boldsymbol0^T & {\mathbf h}_c^T({\mathbf x})\\
\boldsymbol0^T & \cdots & \boldsymbol0^T & {\mathbf h}_{J-1}^T({\mathbf x}) & {\mathbf h}_c^T({\mathbf x})\\
 \boldsymbol0^T & \cdots & \cdots & \boldsymbol0^T & \boldsymbol0^T\\
\end{pmatrix}_{J \times p}
\end{equation}
\normalsize
where ${\mathbf h}^T_j(\cdot) = (h_{j1}(\cdot), \ldots, h_{jp_j}(\cdot))$ and ${\mathbf h}^T_c(\cdot)$ $ = $ $(h_1(\cdot), \ldots, h_{p_c}(\cdot))$ are known predictor functions. We let $\boldsymbol \beta_j$ and $\boldsymbol \zeta$ denote the model parameters associated with ${\mathbf h}^T_j({\mathbf x})$ and ${\mathbf h}^T_c({\mathbf x})$, respectively, then the model parameter vector $\boldsymbol\theta =(\boldsymbol\beta_{1},\boldsymbol\beta_{2},\cdots,\boldsymbol\beta_{J-1},\boldsymbol\zeta)^T \in \mathbb{R}^p$,
and the linear predictor $\boldsymbol\eta_{\mathbf x} = {\mathbf X}_{\mathbf x} \boldsymbol\theta =  (\eta_1^{\mathbf x}, \ldots, \eta_{J-1}^{\mathbf x}, 0)^T \in \mathbb{R}^J$, where $\eta_j^{\mathbf x} = {\mathbf h}_j^T({\mathbf x}) \boldsymbol\beta_j + {\mathbf h}_c^T({\mathbf x}) \boldsymbol\zeta$, $j=1, \ldots, J-1$.

According to Lemmas S.10, S.12 and S.13 in the Supplementary Material of \cite{bu2020}, the categorical probabilities $\boldsymbol\pi_{\mathbf x} = (\pi_1^{\mathbf x}, \ldots, \pi_J^{\mathbf x})^T \in \mathbb{R}^J$ at ${\mathbf x}$ for baseline-category, adjacent-categories and continuation-ratio logit models can be expressed as follows:
\footnotesize
\begin{equation}\label{eq:pi_x_j}
\pi_j^{\mathbf x} = \left\{
\begin{array}{cl}
\frac{\exp\{\eta_j^{\mathbf x}\}}{\exp\{\eta_1^{\mathbf x}\} + \cdots + \exp\{\eta_{J-1}^{\mathbf x}\} + 1} & \mbox{baseline-category}\\
\frac{\exp\{\eta_{J-1}^{\mathbf x} + \cdots +\eta_j^{\mathbf x}\}}{D_j} & \mbox{adjacent-categories}\\
\exp\{\eta_j^{\mathbf x}\} \prod_{l=1}^j (\exp\{\eta_l^{\mathbf x}\} + 1)^{-1} & \mbox{continuation-ratio}
\end{array}\right.
\end{equation}
\normalsize
for $j=1, \ldots, J-1$, where $D_j = \exp\{\eta_{J-1}^{\mathbf x} + \cdots + \eta_1^{\mathbf x}\} + \exp\{\eta_{J-1}^{\mathbf x} + \cdots + \eta_2^{\mathbf x}\} + \cdots +\exp\{\eta_{J-1}^{\mathbf x}\} + 1$, and
\footnotesize
\[
\pi_J^{\mathbf x} = \left\{
\begin{array}{cl}
\frac{1}{\exp\{\eta_1^{\mathbf x}\} + \cdots + \exp\{\eta_{J-1}^{\mathbf x}\} + 1} & \mbox{baseline-category}\\
\frac{1}{D_J} & \mbox{adjacent-categories}\\
\prod_{l=1}^{J-1} (\exp\{\eta_l^{\mathbf x}\} + 1)^{-1} & \mbox{continuation-ratio}
\end{array}\right.
\]
\normalsize
where $D_J = \exp\{\eta_{J-1}^{\mathbf x} + \cdots + \eta_1^{\mathbf x}\} + \exp\{\eta_{J-1}^{\mathbf x} + \cdots + \eta_2^{\mathbf x}\} + \cdots +\exp\{\eta_{J-1}^{\mathbf x}\} + 1$. Note that we provide the expression of $\pi_J^{\mathbf x}$ for completeness while $\pi_J^{\mathbf x} = 1 - \pi_1^{\mathbf x} - \cdots - \pi_{J-1}^{\mathbf x}$ is an easier way for numerical calculations.

As for cumulative logit models, the candidate ${\mathbf x}$ must satisfy $-\infty < \eta_1^{\mathbf x} < \eta_2^{\mathbf x} < \cdots < \eta_{J-1}^{\mathbf x} < \infty$. Otherwise, $0 < \pi_j^{\mathbf x} < 1$ might be violated for some $j=1, \ldots, J$. In other words, the feasible design region should be
\footnotesize
\begin{equation}\label{eq:X_theta_cumul}
{\mathcal X}_{\boldsymbol\theta} = \{{\mathbf x}\in {\mathcal X} \mid -\infty < \eta_1^{\mathbf x} < \eta_2^{\mathbf x} < \cdots < \eta_{J-1}^{\mathbf x} < \infty\}
\end{equation}
\normalsize
which depends on the regression parameter $\boldsymbol\theta$ (see Section~S.14 in the Supplementary Material of  \cite{bu2020} for such an example). For cumulative logit models, if ${\mathbf x} \in {\mathcal X}_{\boldsymbol\theta}$, then 
\footnotesize
\begin{equation}\label{eq:pi_x_j_cumul}
\pi_j^{\mathbf x} = \left\{
\begin{array}{cl}
\frac{\exp\{\eta_1^{\mathbf x}\}}{1+\exp\{\eta_1^{\mathbf x}\}} & j=1\\
\frac{\exp\{\eta_j^{\mathbf x}\}}{1+\exp\{\eta_j^{\mathbf x}\}} - \frac{\exp\{\eta_{j-1}^{\mathbf x}\}}{1+\exp\{\eta_{j-1}^{\mathbf x}\}} & 1 < j < J\\
\frac{1}{1+\exp\{\eta_{J-1}^{\mathbf x}\}} & j=J
\end{array}\right.
\end{equation}
\normalsize
according to Lemma~S.11 of \cite{bu2020}.

Once $\boldsymbol\pi_{\mathbf x} \in \mathbb{R}^J$ is obtained, we can calculate $u_{st}^{\mathbf x} = u_{st}({\boldsymbol\pi}_{\mathbf x})$ based on Theorem~A.2 in \cite{bu2020} as follows:
\begin{itemize}
\footnotesize
	\item[(i)] $u_{st}^{\mathbf x} = u_{ts}^{\mathbf x}$, $s,t=1, \ldots, J$;
	\item[(ii)] $u_{sJ}^{\mathbf x} = 0$ for $s=1, \ldots, J-1$ and $u_{JJ}^{\mathbf x} = 1$;
	\item[(iii)] For $s=1, \ldots, J-1$, $u_{ss}^{\mathbf x}$ is
	\[
		 \left\{\begin{array}{cl}
		\pi_s^{\mathbf x} (1-\pi_s^{\mathbf x}) & \mbox{for baseline-category},\\
		(\gamma_s^{\mathbf x})^2(1-\gamma_s^{\mathbf x})^2((\pi_s^{\mathbf x})^{-1} + (\pi_{s+1}^{\mathbf x})^{-1}) & \mbox{for cumulative},\\
		\gamma_s^{\mathbf x}(1-\gamma_s^{\mathbf x})  & \mbox{for adjacent-categories},\\
		\pi_s^{\mathbf x}(1-\gamma_s^{\mathbf x})(1-\gamma_{s-1}^{\mathbf x})^{-1} & \mbox{for continuation-ratio};
		\end{array}\right.
		\]
		\item[(iv)] For $1\leq s < t \leq J-1$, $u_{st}^{\mathbf x}$ is
		\[
		\left\{\begin{array}{cl}
		-\pi_s^{\mathbf x} \pi_t^{\mathbf x} & \mbox{for baseline-category},\\
		-\gamma_s^{\mathbf x}\gamma_t^{\mathbf x}(1-\gamma_s^{\mathbf x})(1-\gamma_t^{\mathbf x})(\pi_t^{\mathbf x})^{-1} & \mbox{for cumulative}, t-s=1,\\
		0 & \mbox{for cumulative}, t-s>1,\\
		\gamma_s^{\mathbf x}(1-\gamma_t^{\mathbf x}) & \mbox{for adjacent-categories},\\
		0 & \mbox{for continuation-ratio};
		\end{array}\right.
		\]
\normalsize
\end{itemize}
where $\gamma_j^{\mathbf x} = \pi_1^{\mathbf x} + \cdots + \pi_j^{\mathbf x}$, $j=1, \ldots, J-1$; $\gamma_0^{\mathbf x}\equiv 0$ and $\gamma_J^{\mathbf x}\equiv 1$.

\section{Example that \texorpdfstring{${\mathbf F}_{{\mathbf x}} = {\mathbf F}_{{\mathbf x}'}$}{Lg} with \texorpdfstring{${\mathbf x} \neq {\mathbf x}'$}{Lg}}\label{ex:F_x_1_x_2}

Consider a special MLM~\eqref{logitunifiedmodel} with proportional odds (po) (see Section~S.7 in the Supplementary Material of \cite{bu2020} for more technical details). Suppose $d=2$ and a feasible design point ${\mathbf x} = (x_1, x_2)^T \in [a, b]\times [-c, c] = {\cal X}$, $c > 0$, $J\geq 2$, ${\mathbf h}_c({\mathbf x}) = (x_1, x_2^2)^T$. Then the model matrix at ${\mathbf x} = (x_1, x_2)^T$ is 
\[
{\mathbf X}_{\mathbf x} = \left(
\begin{array}{cccccc}
1 & 0 & \cdots & 0 & x_1 & x_2^2\\
0 & 1 & \ddots & \vdots & \vdots\\
\vdots & \ddots & \ddots & 0 & x_1 & x_2^2\\
0 & \cdots & 0 & 1 & x_1 & x_2^2\\
0 & \cdots & 0 & 0 & 0 & 0
\end{array}\right)_{J\times (J+1)}
\]
Then $p=J+1$. Let $\boldsymbol\theta = (\beta_1, \ldots, \beta_{J-1}, \zeta_1, \zeta_2)^T \in \mathbb{R}^{J+1}$ be the model parameters (since $\boldsymbol\theta$ is fixed, we may assume that ${\cal X} = {\cal X}_{\boldsymbol\theta}$ if the model is a cumulative logit model). Let ${\mathbf x}' = (x_1, -x_2)^T$. Then ${\mathbf X}_{\mathbf x} = {\mathbf X}_{{\mathbf x}'}$ and thus $\boldsymbol\eta_{\mathbf x} = \boldsymbol\eta_{{\mathbf x}'}$. According to \eqref{eq:pi_x_j} (or \eqref{eq:pi_x_j_cumul}), we obtain $\boldsymbol\pi_{\mathbf x} = \boldsymbol\pi_{{\mathbf x}'}$ and then ${\mathbf U}_{\mathbf x} = {\mathbf U}_{{\mathbf x}'}$~. The Fisher information matrix at ${\mathbf x}$ is ${\mathbf F}_{\mathbf x} = {\mathbf X}_{\mathbf x}^T {\mathbf U}_{\mathbf x} {\mathbf X}_{\mathbf x} = {\mathbf X}_{{\mathbf x}'}^T {\mathbf U}_{{\mathbf x}'} {\mathbf X}_{{\mathbf x}'} = {\mathbf F}_{{\mathbf x}'}$. Note that ${\mathbf x} \neq {\mathbf x}'$ if $x_2 \neq 0$.

\section{First-order derivative of sensitivity function}\label{sec:MLM_first_order}

As mentioned in Section~\ref{sec:F_x_for_MLM}, to apply Algorithm~\ref{algo:ForLion_general} for MLM, we need to calculate the first-order derivative of the sensitivity function $d({\mathbf x}, \boldsymbol{\xi})$.

Recall that the first $k$ ($1\leq k\leq d$) factors are continuous. Given ${\mathbf x} = (x_1, \ldots, x_d)^T \in {\cal X}$, for each $i=1, \ldots, k$, according to Formulae~17.1(a), 17.2(a) and 17.7 in \cite{seber2008}, 
\footnotesize
\begin{eqnarray}
& & \frac{\partial d({\mathbf x}, \boldsymbol\xi)}{\partial x_i}\ =\ \frac{\partial {\rm tr}({\mathbf F}(\boldsymbol\xi)^{-1} {\mathbf F}_{\mathbf x})}{\partial x_i} \nonumber \\
&=& {\rm tr}\left( {\mathbf F}(\boldsymbol\xi)^{-1} \frac{\partial {\mathbf F}_{\mathbf x}}{\partial x_i}\right)\nonumber\\
&=& {\rm tr}\left( {\mathbf F}(\boldsymbol\xi)^{-1} \left[\frac{\partial {\mathbf X}_{\mathbf x}^T}{\partial x_i} {\mathbf U}_{\mathbf x} {\mathbf X}_{\mathbf x} + {\mathbf X}_{\mathbf x}^T \frac{\partial {\mathbf U}_{\mathbf x}}{\partial x_i} {\mathbf X}_{\mathbf x}\right.\right.\nonumber\\
&+& \left.\left. {\mathbf X}_{\mathbf x}^T {\mathbf U}_{\mathbf x} \frac{\partial {\mathbf X}_{\mathbf x}}{\partial x_i}\right]\right) \label{eq:partial_d_partial_x_i}
\end{eqnarray}
\normalsize
where
\footnotesize
\begin{equation}\label{eq:p_X_x_p_x_i}
\frac{\partial {\mathbf X}_{\mathbf x}}{\partial x_i}= \begin{pmatrix}
 \frac{\partial {\mathbf h}_1^T({\mathbf x})}{\partial x_i} &  \boldsymbol0^T & \cdots & \boldsymbol0^T& \frac{\partial {\mathbf h}_c^T({\mathbf x})}{\partial x_i}\\
 \boldsymbol0^T &  \frac{\partial {\mathbf h}_2^T({\mathbf x})}{\partial x_i} &\ddots & \vdots & \vdots\\
\vdots &  \ddots& \ddots &  \boldsymbol0^T & \frac{\partial {\mathbf h}_c^T({\mathbf x})}{\partial x_i}\\
\boldsymbol0^T & \cdots & \boldsymbol0^T & \frac{\partial {\mathbf h}_{J-1}^T({\mathbf x})}{\partial x_i} & \frac{\partial {\mathbf h}_c^T({\mathbf x})}{\partial x_i}\\
 \boldsymbol0^T & \cdots & \cdots & \boldsymbol0^T & \boldsymbol0^T\\
\end{pmatrix}_{J \times p}
\end{equation}
\normalsize
$\frac{\partial {\mathbf U}_{\mathbf x}}{\partial x_i} = \left(\frac{\partial u^{\mathbf x}_{st}}{\partial x_i}\right)_{s,t=1, \ldots, J}$ with 
\begin{eqnarray}
    \frac{\partial u^{\mathbf x}_{st}}{\partial x_i} &=& \frac{\partial u^{\mathbf x}_{st}}{\partial \boldsymbol{\pi}_{\mathbf x}^T} \cdot \frac{\partial \boldsymbol{\pi}_{\mathbf x}}{\partial \boldsymbol{\eta}_{\mathbf x}^T} \cdot \frac{\partial \boldsymbol{\eta}_{\mathbf x}}{\partial x_i}\nonumber \\
	&=& \frac{\partial u^{\mathbf x}_{st}}{\partial \boldsymbol{\pi}_{\mathbf x}^T} \cdot \left({\mathbf C}^T {\mathbf D}_{\mathbf x}^{-1} {\mathbf L}\right)^{-1} \cdot \frac{\partial {\mathbf X}_{\mathbf x}}{\partial x_i} \cdot \boldsymbol{\theta} \label{eq:p_u_st_p_x_i}
\end{eqnarray}
${\mathbf C}$ and ${\mathbf L}$ defined as in \eqref{logitunifiedmodel}, and ${\mathbf D}_{\mathbf x} = {\rm diag}({\mathbf L} \boldsymbol{\pi}_{\mathbf x})$. Explicit formula of $({\mathbf C}^T {\mathbf D}_{\mathbf x}^{-1} {\mathbf L})^{-1}$ can be found in  Section~S.3 in the Supplementary Material of \cite{bu2020} with ${\mathbf x}_i$ replaced by $ {\mathbf x}$.  As for $\frac{\partial u^{\mathbf x}_{st}}{\partial \boldsymbol{\pi}_{\mathbf x}^T}$, we have the following explicit formulae
\begin{itemize}
	\item[(i)] $\frac{\partial u_{st}^{\mathbf x}}{\partial \boldsymbol{\pi}_{\mathbf x}} = \frac{\partial u_{ts}^{\mathbf x}}{\partial \boldsymbol{\pi}_{\mathbf x}}$, $s,t=1, \ldots, J$;
	\item[(ii)] $\frac{\partial u_{sJ}^{\mathbf x}}{\partial \boldsymbol{\pi}_{\mathbf x}} = {\mathbf 0} \in \mathbb{R}^J$ for $s=1, \ldots, J$;
	\item[(iii)] For $s=1, \ldots, J-1$, $\frac{\partial u_{ss}^{\mathbf x}}{\partial \boldsymbol{\pi}_{\mathbf x}}$ is
\footnotesize
	\[
	\left\{\begin{array}{cl}
		\left(\pi_s^{\mathbf x} {\mathbf 1}_{s-1}^T, 1-\pi_s^{\mathbf x}, \pi_s^{\mathbf x} {\mathbf 1}_{J-s}^T\right)^T & \mbox{for baseline-category}\\
		u_{ss}^{\mathbf x} \left[\left(\frac{2}{\gamma_s^{\mathbf x}} {\mathbf 1}_s^T, \frac{2}{1-\gamma_s^{\mathbf x}} {\mathbf 1}_{J-s}^T\right)^T\right. & \\
		\left. - \frac{\pi_{s+1}^{\mathbf x} {\mathbf e}_s}{\pi_s^{\mathbf x} (\pi_s^{\mathbf x} + \pi_{s+1}^{\mathbf x})} - \frac{\pi_s^{\mathbf x} {\mathbf e}_{s+1}}{\pi_{s+1}^{\mathbf x} (\pi_s^{\mathbf x} + \pi_{s+1}^{\mathbf x})}\right] & \mbox{for cumulative}\\
		\left( (1-\gamma_s^{\mathbf x}) {\mathbf 1}_s^T, \gamma_s^{\mathbf x} {\mathbf 1}_{J-s}^T\right)^T & \mbox{for adjacent-categories}\\
		\left({\mathbf 0}_{s-1}^T, \frac{(1-\gamma_s^{\mathbf x})^2}{(1-\gamma_{s-1}^{\mathbf x})^2}, \frac{(\pi_s^{\mathbf x})^2 {\mathbf 1}_{J-s}^T}{(1-\gamma_{s-1}^{\mathbf x})^2}\right)^T & \mbox{for continuation-ratio}
	\end{array}\right.
	\]
 \normalsize
	where ${\mathbf e}_s$ is the $J\times 1$ vector with the $s$th coordinate $1$ and all others $0$, ${\mathbf 1}_s$ is the $s\times 1$ vector of all $1$, and ${\mathbf 0}_s$ is the $s\times 1$ vector of all $0$.
	\item[(iv)] For $1\leq s < t \leq J-1$, $\frac{\partial u_{st}^{\mathbf x}}{\partial \boldsymbol{\pi}_{\mathbf x}}$ is
 \footnotesize
	\[
	\left\{\begin{array}{cl}
		\left({\mathbf 0}_{s-1}^T, -\pi_t^{\mathbf x}, {\mathbf 0}_{t-s-1}^T, -\pi_s^{\mathbf x}, {\mathbf 0}_{J-t}^T\right)^T & \mbox{for baseline-category}\\
		\left(-(1-\gamma_s^{\mathbf x})(1-\gamma_t^{\mathbf x})\left(1 + \frac{2\gamma_s^{\mathbf x}}{\pi_t^{\mathbf x}}\right) {\mathbf 1}_s^T, \right. & \\ -\gamma_s^{\mathbf x} (1-\gamma_t^{\mathbf x}) \left[1 - \frac{\gamma_s^{\mathbf x} (1-\gamma_t^{\mathbf x})}{(\pi_t^{\mathbf x})^2}\right], & \\
		\left. -\gamma_s^{\mathbf x} \gamma_t^{\mathbf x} \left[ 1 + \frac{2(1-\gamma_t^{\mathbf x})}{\pi_t}\right] {\mathbf 1}_{J-s-1}^T\right)^T & \mbox{for cumulative}, t-s=1\\
		{\mathbf 0}_J & \mbox{for cumulative}, t-s>1\\
		\left( (1-\gamma_t^{\mathbf x}) {\mathbf 1}_s^T, {\mathbf 0}_{t-s}^T, \gamma_s^{\mathbf x} {\mathbf 1}_{J-t}^T\right)^T & \mbox{for adjacent-categories}\\
		{\mathbf 0}_J & \mbox{for continuation-ratio}
	\end{array}\right.
	\]
\normalsize
\end{itemize}
where $\gamma_j^{\mathbf x} = \pi_1^{\mathbf x} + \cdots + \pi_j^{\mathbf x}$, $j=1, \ldots, J-1$; $\gamma_0^{\mathbf x}\equiv 0$ and $\gamma_J^{\mathbf x}\equiv 1$.

Thus the explicit formulae for $\frac{\partial d({\mathbf x}, \boldsymbol\xi)}{\partial x_i}$, $i=1, \ldots, k$ can be obtained via \eqref{eq:partial_d_partial_x_i}. Only $\frac{\partial {\mathbf X}_{\mathbf x}}{\partial x_i}$ is related to $i$, which may speed up the computations.

\end{appendices}




\clearpage
\newpage
\begin{center}
{\Large\bf ForLion: A New Algorithm for
D-optimal Designs under
General Parametric Statistical Models
with Mixed Factors}
\vspace{.25cm}

{\large{Yifei Huang$^1$,  Keren Li$^2$, Abhyuday Mandal$^3$, and Jie Yang$^{1*}$\footnote{*CONTACT jyang06@uic.edu }
\hspace{.2cm}}}
\vspace{.2cm}

{$^1$ University of Illinois at Chicago}

{$^2$ University of Alabama at Birmingham}

{$^3$ University of Georgia}
\end{center}

\vspace{.55cm}
 \centerline{\bf Supplementary Material}
\vspace{.55cm}
\fontsize{9}{11.5pt plus.8pt minus .6pt}\selectfont
\noindent
{\bf S.1 Commonly used GLMs:} A list of commonly used GLM models, corresponding link functions, $\nu$ functions, and their first-order derivatives;\\
{\bf S.2 Technical details of house flies example:} Technical details of applying the ForLion algorithm to the emergence of house flies example;\\
{\bf S.3 Example: Minimizing surface defects:} An example with cumulative logit po model that shows the advantages of the ForLion algorithm;\\
{\bf S.4 Fisher information matrix for GLMs:} Formulae for computing Fisher information matrix for GLMs;\\
{\bf S.5 First-order derivative of sensitivity function for GLMs:} Formulae of $\partial d({\mathbf x}, \boldsymbol\xi)/\partial x_i$ for GLMs;\\
{\bf S.6 GLMs with main-effects continuous factors:} Details of GLMs with main-effects continuous factors;\\
{\bf S.7 Electrostatic discharge example supplementary:} The optimal design table for electrostatic discharge example and a simulation study on the effects of merging threshold $\delta$;\\
{\bf S.8 Assumptions needed for Theorem~\ref{thm:doptimality}};\\
{\bf S.9 Proofs:} Proofs for theorems in this paper.

\par

\setcounter{section}{0}
\setcounter{equation}{0}
\setcounter{table}{0}
\setcounter{figure}{0}
\def\thesection{S.\arabic{section}}
\def\theequation{S\arabic{section}.\arabic{equation}}
\def\thetable{S\arabic{table}}
\def\thefigure{S\arabic{figure}}

\fontsize{9}{11.5pt plus.8pt minus .6pt}\selectfont

\section{Commonly used GLMs}\label{subsec:glm.formula}

In this section, we provide the formulae of the link function, $\nu(\eta)$, and $\nu'(\eta)$ for commonly used GLMs, which are needed for the ForLion algorithm (see Section~\ref{sec:GLM:forlion}, as well as Sections~\ref{sec:GLM_Fisher}, \ref{sec:GLM_sensitivity_derivative}, and \ref{sec:maineffects_supp} in the Supplementary Material).

\begin{enumerate}
\item Bernoulli($\mu$) with logit link\\
	Link function:\\ $\eta=g(\mu) = \log(\mu/(1-\mu))$;\\
	$\nu(\eta) = e^\eta/(1+e^\eta)^2$;\\
	$\nu'(\eta) = -e^\eta (e^\eta - 1)/(1 + e^\eta)^3$.
\item Bernoulli($\mu$) with probit link\\
	Link function: $\eta=g(\mu) = \Phi^{-1}(\mu)$;\\
	$\nu(\eta) = \phi(\eta)^2/\left\{\Phi(\eta)[1-\Phi(\eta)]\right\}$;
	\begin{eqnarray*}
			\nu'(\eta) &=& -\frac{2\eta \phi^2(\eta)}{[1-\Phi(\eta)]\Phi(\eta)} + \frac{\phi^3(\eta) [2\Phi(\eta) -1]}{[1-\Phi(\eta)]^2 \Phi^2(\eta)}
	\end{eqnarray*}
	Note that $\phi'(\eta)=-\eta \phi(\eta)$ and $\phi''(\eta)=(\eta^2-1)\phi(\eta)$.

\item Bernoulli($\mu$) with cloglog link\\
Link function:\\ $\eta=g(\mu) = \log(-\log(1-\mu))$;\\
	$\nu(\eta) = e^{2\eta}/\left(\exp\{e^\eta\}-1\right)$;
	\begin{eqnarray*}
			\nu'(\eta) &=& \frac{2e^{2\eta}}{\exp\{e^\eta\}-1} -\frac{\exp\{3\eta + e^\eta\}}{(\exp\{e^\eta\}-1)^2}
	\end{eqnarray*}
	
\item Bernoulli($\mu$) with loglog link\\
Link function:\\ $\eta=g(\mu) = \log(-\log\mu)$;\\
    In this case, $\nu(\eta)$, $\nu'(\eta)$,
    are exactly the same as Bernoulli($\mu$) with cloglog link.\\
	An alternative (strictly increase) loglog link \citep{ytm2016} is $\eta = g(\mu) = -\log(-\log\mu)$. Then $\nu(\eta)=\exp\{-2\eta\}/[\exp\{e^{-\eta}\}-1]$.

\item Bernoulli($\mu$) with cauchit link\\
	Link function:\\ $\eta=g(\mu) = {\rm tan}(\pi (\mu-1/2))$;\\
	$\nu(\eta) = (1+\eta^2)^{-2}[\pi^2/4 - {\rm arctan}^2(\eta)]^{-1}$;
	\begin{eqnarray*}
		\nu'(\eta) &=& \frac{16}{(1+\eta^2)^3}\left(\frac{-\eta}{\pi^2 -4 {\rm arctan}^2(\eta)}\right.\\
  &+& \left.\frac{2 {\rm arctan}(\eta)}{[\pi^2 -4 {\rm arctan}^2(\eta)]^2}\right)
	\end{eqnarray*}

\item Bernoulli($\mu$) with $t$ link \citep{liu2004}\\
	Link function: $\eta=g(\mu) = F_\nu^{-1} (\mu)$, where $F_\nu$ is the cumulative distribution function of $t$ distribution with degrees of freedom $\nu>0$;\\
	$\left(g^{-1}\right)'(\eta) = f_\nu (\eta) = \frac{\Gamma\left(\frac{\nu+1}{2}\right)}{\sqrt{\nu \pi}\ \Gamma\left(\frac{\nu}{2}\right)} \left(1 + \frac{\eta^2}{\nu}\right)^{-\frac{\nu+1}{2}}$, where $f_\nu$ is the probability density function of $t$ distribution with degrees of freedom $\nu>0$. Then
	\begin{eqnarray*}
	    \nu(\eta) &=& \frac{f^2_\nu(\eta)}{F_\nu(\eta) \left[1-F_\nu(\eta)\right]}\\
\nu'(\eta) &=& -\frac{2\eta f_\nu^2(\eta)}{[1-F_\nu(\eta)]F_\nu(\eta)} \\
&+& \frac{f_\nu^3(\eta) [2F_\nu(\eta) -1]}{[1-F_\nu(\eta)]^2 F_\nu^2(\eta)}
	\end{eqnarray*}

\item Poisson($\mu$) with log link\\
	Link function: $\eta=g(\mu) = \log(\mu)$;\\
	$\nu(\eta) = \nu'(\eta) = e^\eta$. 

\item Gamma($k$, $\mu/k$) with reciprocal link, $k >0$ is known\\
	Link function: $\eta=g(\mu) = \mu^{-1}$, $\mu>0$;\\
	$\nu(\eta) = k\eta^{-2}$;\\
	$\nu'(\eta) = -2k\eta^{-3}$.

\item Normal($\mu$, $\sigma^2$) with identity link, that is, the usual linear regression model\\
	Link function: $\eta=g(\mu) = \mu$;\\
	$\nu(\eta) \equiv \sigma^{-2}$; $\nu'(\eta) = 0$.
\item Inverse Gaussian($\mu$, $\lambda$) with inverse squared link, $\lambda >0$ is known\\
	Link function: $\eta=g(\mu) = \mu^{-2}$;\\
	$\nu(\eta) = \frac{\lambda}{4} \eta^{-3/2}$;\\
	$\nu'(\eta) = -\frac{3}{8}\lambda \eta^{-5/2}$.
\end{enumerate}

\par

\section{Technical details of house flies example}\label{sec: MLM_fly_techdetail}

In this section, we follow Section~\ref{sec:MLM_example_fly} and provide more technical details of the ForLion algorithm for the emergence of the house flies example. 
To use our ForLion algorithm, $p_1=3$, $p_2=2$, $p_c=0$, and $p=p_1+p_2+p_c=5$. Given ${\mathbf x}=x \in {\mathcal X} = [80, 200]$ or $[0, 200]$, we have ${\mathbf h}_1({\mathbf x}) = (1, x, x^2)^T$, ${\mathbf h}_2({\mathbf x}) = (1,x)^T$, ${\mathbf h}_c({\mathbf x}) = 0$. Correspondingly, 
\[
{\mathbf X}_{\mathbf x} = \left(
\begin{array}{ccccc}
1 & x & x^2 & 0 & 0\\
0 & 0  & 0 & 1 & x\\
0 & 0 & 0 & 0 & 0
\end{array}
\right)_{3\times 5}
\]
\[
\frac{\partial {\mathbf X}_{\mathbf x}}{\partial x} = \left(
\begin{array}{ccccc}
0 & 1 & 2x & 0 & 0\\
0 & 0  & 0 & 0 & 1\\
0 & 0 & 0 & 0 & 0
\end{array}
\right)_{3\times 5}
\]
and $\boldsymbol\eta^{\mathbf x} = {\mathbf X}_{\mathbf x} \boldsymbol\theta = (\beta_{11} + \beta_{12} x + \beta_{13} x^2, \beta_{21} + \beta_{22} x, 0)^T$.
To calculate $\mathbf F_{\mathbf x} = {{\mathbf X}_{\mathbf x}}^T {\mathbf U}_{\mathbf x} {\mathbf X}_{\mathbf x}$, we also need 
\[
{\mathbf U}_{\mathbf x}=
\left(
\begin{array}{ccc}
\frac{\pi^{\mathbf x}_{1} (1-\gamma_1^{\mathbf x})}{1-\gamma_0^{\mathbf x}} & 0 & 0\\
0 & \frac{\pi^{\mathbf x}_{2} (1-\gamma_2^{\mathbf x})}{1-\gamma_1^x} & 0\\
0 & 0 & 1\\
\end{array}
\right)_{3\times3}
\]
with $\gamma^{\mathbf x}_0 = 0$, $\gamma^{\mathbf x}_1 = \pi_1^{\mathbf x}$, $\gamma^{\mathbf x}_2 = \pi_1^{\mathbf x} + \pi_2^{\mathbf x}$.

For Step~$5^\circ$ of the ForLion algorithm, we have $k=d=1$. To use the R function $\tt{optim}$ with option``L-BFGS-B'' (a quasi-Newton method), we need to calculate $\frac{\partial d({\mathbf x}, {\boldsymbol\xi}_t)}{\partial x}$ based on \eqref{eq:partial_d_partial_x_i} and \eqref{eq:p_u_st_p_x_i}, where
$\frac{\partial u^{\mathbf x}_{s3}}{\partial \boldsymbol \pi_{\mathbf x}^T} = \frac{\partial u^{\mathbf x}_{3s}}{\partial \boldsymbol \pi_{\mathbf x}^T}=(0, 0, 0)$ for $s = 1, 2, 3$; 
$\frac{\partial u^{\mathbf x}_{11}}{\partial \boldsymbol \pi_{\mathbf x}^T} = ((1-\gamma_1^{\mathbf x})^2, (\pi^{\mathbf x}_{1})^2, (\pi^{\mathbf x}_{1})^2)$; 
$\frac{\partial u^{\mathbf x}_{22}}{\partial \boldsymbol \pi_{\mathbf x}^T} = (0, \frac{(1-\gamma_2^{\mathbf x})^2}{(1-\gamma_1^{\mathbf x})^2},  \frac{(\pi^{\mathbf x}_{2})^2}{(1-\gamma_1^{\mathbf x})^2})$;
$\frac{\partial u^{\mathbf x}_{12}}{\partial \boldsymbol \pi_{\mathbf x}^T} = \frac{\partial u^{\mathbf x}_{21}}{\partial \boldsymbol \pi_{\mathbf x}^T} = (0, 0, 0)$; and 
\[
(\mathbf C^T \mathbf D_{\mathbf x}^{-1} \mathbf L)^{-1} = 
\left(
\begin{array}{ccc}
    \pi^{\mathbf x}_{1}(1-\gamma_1^{\mathbf x}) & 0 & \pi^{\mathbf x}_{1} \\
     -\pi^{\mathbf x}_{1}\pi^{\mathbf x}_{2} & \frac{\pi^{\mathbf x}_{2}(1-\gamma_2^{\mathbf x})}{1-\gamma_1^{\mathbf x}} & \pi^{\mathbf x}_{2}\\
     -\pi^{\mathbf x}_{1}\pi^{\mathbf x}_{3} & -\frac{\pi^{\mathbf x}_{2}\pi^{\mathbf x}_{3}}{1-\gamma_1^{\mathbf x}} & \pi^{\mathbf x}_{3}
\end{array}
\right)_{3 \times 3}
\]

\section{Example: Minimizing surface defects }\label{sec:MLM_example_surface}

In Chapter $4$ of \cite{phadke1989}, a polysilicon deposition process for manufacturing very large-scale integrated (VLSI) circuits was described with six control factors, namely, Cleaning Method, Deposition temperature ($^\circ$C), Deposition pressure (mtorr), Nitrogen flow rate (seem), Silane flow rate (seem), and Settling time (minutes). \cite{wu2008} used the relevant experiment as an illustrative example and categorized the response Surface Defects from a count number to one of the five ordered categories, namely, Practically no surface defects (I, $0\sim 3$), Very few defects (II, $4\sim 30$), Some defects (III, $31\sim 300$), Many defects (IV, $301\sim 1000$), and Too many defects (V, $1001$ and above). \cite{lukemire2022} utilized this experiment as an example under a cumulative logit proportional odds (po) model. The factor Cleaning Method is binary and the other five control factors are continuous ones.
The levels and ranges of these factors are listed in Table~\ref{tab:example_surface_levels} (see also Table~6 in \cite{lukemire2022}). 

\begin{table*}[hbt!]
\caption{Factors/parameters in the example of minimizing surface defects}
    \label{tab:example_surface_levels}
    \centering
    \begin{tabular}{c|c c c}\hline
    Factor& Factor & Parameter & Parameter\\
        (Unit, Parameter) & Levels/Range & Nominal Value & Prior Distribution \\ \hline
        Cleaning Method ($\beta_1$) & CM$_2$, CM$_3$ & $-0.970$ & U$(-1,0)$ \\
        Deposition temperature ($^\circ$C,  $\beta_2$) & $[-25, 25]$ & $0.077$ & U$(0, 0.2)$ \\
        Deposition pressure (mtorr,  $\beta_3$) &  $[-200, 200]$ & $0.008$ & U$(-0.1,0.1)$ \\
        Nitrogen flow (seem, $\beta_4$) & $[-150, 0]$ & $-0.007$ & U$(-0.1,0.1)$ \\
        Silane flow (seem, $\beta_5$) &  $[-100, 0]$ & $0.007$ & U$(-0.1,0.1)$ \\
        Settling time (min, $\beta_6$) &  $[0, 16]$ & $0.056$ & U$(0, 0.2)$\\
        Cut point $1$ ($\theta_1$) & (Intercept) & $-1.113$ & U$(-2,-1)$\\
        Cut point $2$ ($\theta_2$) & (Intercept) & $0.183$ & U$(-0.5, 0.5)$\\
        Cut point $3$ ($\theta_3$) & (Intercept) & $1.518$ & U$(1, 2)$\\
        Cut point $4$ ($\theta_4$) & (Intercept) & $2.639$ & U$(2.5, 3.5)$\\\hline
    \end{tabular}
\end{table*}

In our notations, the cumulative logit model used by \cite{lukemire2022} for the surface defects experiment can be expressed as follows:
\footnotesize
\begin{eqnarray}
& & \log\left(\frac{\pi_{i1}+\cdots+\pi_{ij}}{\pi_{i,j+1}+\cdots+\pi_{iJ}}\right)\nonumber\\
&=& \theta_{j} - \beta_1 x_{i1} - \beta_2 x_{i2} - \beta_3 x_{i3}\nonumber\\
&-& \beta_4 x_{i4} - \beta_5 x_{i5} - \beta_6 x_{i6}
\label{eq:surface_cumulative}
\end{eqnarray}
\normalsize
with $i=1, \ldots, m$ and $j=1, 2, 3, 4$.

In this example, $d=6$, $k=5$, $p=10$, and $\boldsymbol\theta = (\theta_1, \theta_2, \theta_3, \theta_4, \beta_1, \beta_2, \beta_3, \beta_4, \beta_5, \beta_6)^T$. Given $\mathbf x = (x_1, x_2, x_3, x_4, x_5, x_6)^T$, ${\mathbf h}_1^T({\mathbf x}) = \cdots = {\mathbf h}_4^T({\mathbf x}) \equiv 1$, ${\mathbf h}_c^T({\mathbf x}) \equiv {\mathbf x}$.

\begin{figure*}[t!] 
     \centering
     \begin{minipage}[t]{0.45\textwidth}
         \centering
         \includegraphics[height=2.3in]{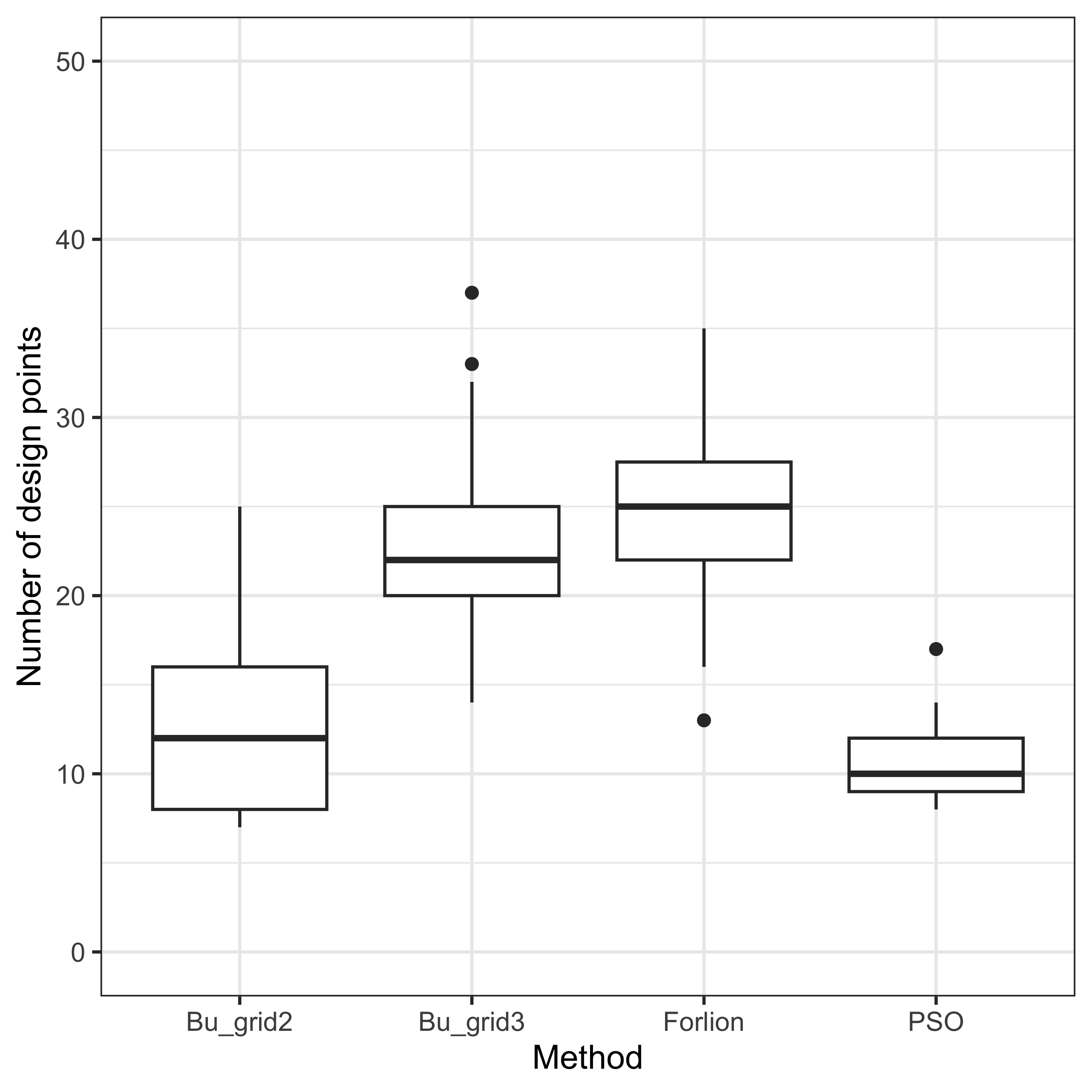}

         (a) Number of design points
     \end{minipage}%
     ~ 
     \begin{minipage}[t]{0.45\textwidth}
         \centering
         \includegraphics[height=2.3in]{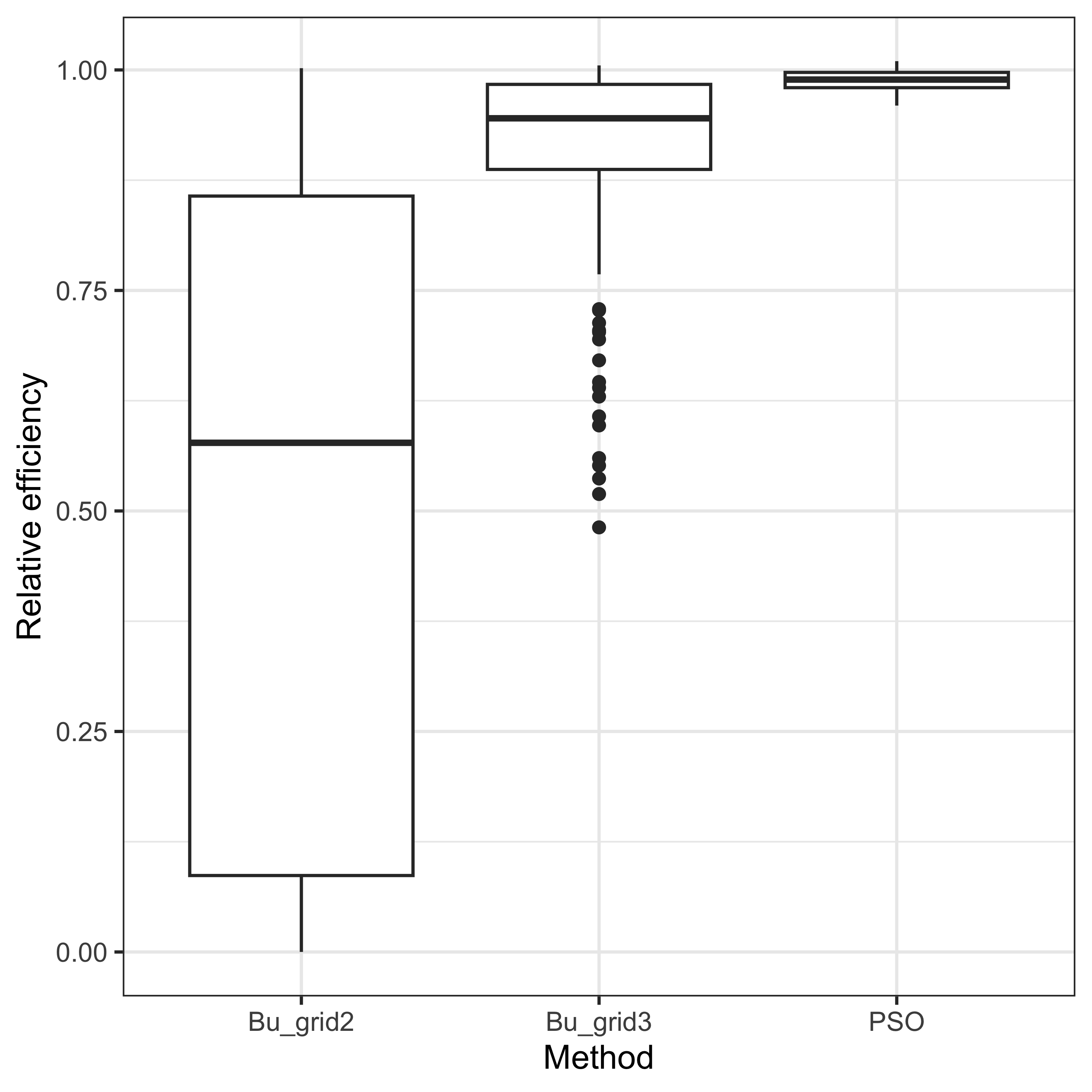}

         (b) Relative efficiency w.r.t. ForLion
     \end{minipage}
     \caption{Boxplots of 100 simulations for Bu's, ForLion, and PSO algorithms on the surface defects study with five categories}\label{fig:mlm_surface}
\end{figure*}

We compare the performance of the PSO algorithm described in \cite{lukemire2022}, \cite{bu2020}'s lift-one algorithm on grid points, and our ForLion algorithm by a simulation study. For the PSO algorithm, following \cite{lukemire2022}, we use $24$ support points and $20$ particles with up to $50,000$ maximum iterations.
For \cite{bu2020}'s lift-one algorithm, we use grid lengths $2$ and $3$, respectively, for discretizing the five continuous factors, which means that we use $2$ or $3$ equally spaced points within the ranges of each continuous factor to obtain the grid points. The corresponding design space consists of $2\times 2^5=64$ or $2\times 3^5=486$ design points, respectively. Grids with a length of $4$ or more are skipped in this case due to extremely long running time. For the ForLion algorithm, we use merging threshold $\delta = 0.2$ and converging threshold $\epsilon=10^{-6}$ for illustration. Both ForLion and Bu's algorithms are written in R, while PSO's is implemented in Julia. We randomly simulate $100$ parameter vectors from the uniform distributions listed in Table~\ref{tab:example_surface_levels} as the assumed true values, respectively. Figure~\ref{fig:mlm_surface} shows the numbers of design points of Bu's Grid-2, Bu's Grid-3, ForLion's, and PSO's designs and their relative efficiencies with respect to the ForLion designs. The median numbers of the design points of the constructed designs are $10$ for PSO, $12$ for Bu's Grid-2, $22$ for Bu's Grid-3, and $25$ for ForLion. The average of relative efficiencies compared to ForLion's, defined as $[f(\boldsymbol{\xi})/f(\boldsymbol\xi_{\rm ForLion})]^{1/10}$, are $49.12\%$ for Bu's Grid-2,  $89.38\%$ for Bu's Grid-3, and $98.82\%$ for PSO. The median running time on the desktop described in Section~\ref{sec:MLM_example_fly} is $0.95$s for Bu's Grid-2, $433.8$s for Bu's Grid-3, $25.50$s for PSO, and $146.83$s for ForLion.

As a conclusion, \cite{bu2020}'s lift-one algorithm with grid points does not work well for this case with five continuous factors. Neither Grid-2 nor Grid-3 has satisfactory relative efficiencies, while Grid-4 or above is computationally too heavy. PSO algorithm, in this case, has again the advantages of reduced computational time and numbers of design points. 
The advantage of the ForLion algorithm is that its relative efficiencies are apparently higher than PSO's, although the improvement is not as impressive as in the electrostatic discharge experiment (see Example~\ref{ex:ESD}).

To check whether the improvement of ForLion's designs against PSO's in terms of relative efficiency is significant, we test if the mean relative efficiency is significantly less than $1$. Supported by the Box-Cox Transformations \citep{box1964analysis,venables2002modern}, we run a one-sided $t$-test to check if $f(\boldsymbol{\xi}_{\rm PSO})/f(\boldsymbol\xi_{\rm ForLion})$ is equal to or less than $1$. The $p$-value $< 2.2\times 10^{-16}$ suggests that the ratio is significantly less than $1$. Actually, a $95\%$ confidence interval for $f(\boldsymbol{\xi}_{\rm PSO})/f(\boldsymbol\xi_{\rm ForLion})$ is $(0.873, 0.915)$. That is, from the optimization point of view, the maximum value of the objective function $f(\boldsymbol{\xi})$ achieved by PSO is only about $90\%$ on average of the one achieved by ForLion in this case.

\par

\section{Fisher information matrix for GLMs}\label{sec:GLM_Fisher}

In this section, we provide the details of the Fisher information matrix for the GLM~\eqref{eq:glm}. The number of covariates $d$ could be much less than the number of parameters $p$. For example, for an experiment with $d=3$ factors $x_1, x_2, x_3$, the model $g(E(Y))=\beta_0 + \beta_1 x_1 + \beta_2 x_2 + \beta_3 x_3 + \beta_4 x_1 x_2$ contains $p=5$ parameters which covers the intercept and an interaction term $x_1x_2$~.

Assuming that there are $m$ distinct covariate combinations ${\bf x}_1, \ldots, {\bf x}_m$ with the numbers of replicates $n_1, \ldots, n_m$, respectively, where $n_1 + \cdots + n_m = n$. The Fisher information matrix ${\mathbf F}$ can be written as \citep{ym2015}
\begin{equation}\label{eq:Fisher_GLM}
{\mathbf F}=n{\mathbf X}^T{\mathbf W}{\mathbf X}=n \sum_{i=1}^m w_i \nu_i {\bf X}_i {\bf X}_i^T
\end{equation}
where ${\mathbf X}=({\mathbf X}_1, \ldots, {\mathbf X}_m)^T$ is an $m\times p$ matrix, known as the model matrix, and ${\mathbf W}={\rm diag}\{w_1\nu_1,\ldots, w_m\nu_m\}$ is an $m\times m$ diagonal matrix with $w_i = n_i/n$ and
$$\nu_i = \frac{1}{{\rm Var}(Y_i)}\left(\frac{\partial \mu_i}{\partial \eta_i}\right)^2$$
Here $w_i$ is the proportion of experimental units assigned to the design point ${\mathbf x}_i$ and $\nu_i$ represents how much information the design point ${\mathbf x}_i$ could provide.

For typical applications, the link function $g$ is one-to-one and differentiable. Following \cite{ym2015}, we assume that $E(Y)$ itself determines ${\rm Var}(Y)$. Then there exists a function $s$ such that ${\rm Var}(Y_i) = s(\eta_i)$. Let $\nu =  \left[\left(g^{-1}\right)'\right]^2/s$. Then $\nu_i = \nu(\eta_i) = \nu\left({\mathbf X_i}^T\boldsymbol\beta\right)$ for each $i=1, \ldots, m$. 
Please see Table~5 in the Supplementary Material of \cite{huang2023constrained} for some examples of the function $\nu$ (see also \cite{pmcc1989, christensen2015analysis, ym2015}). Besides, according to \cite{liu2004}, the $t$ link family $g(\mu) = F_\nu^{-1}(\mu)$ incorporates logit and probit links approximately, where $F_\nu$ is the cumulative distribution function of $t$ distribution with degrees of freedom $\nu > 0$. In Section~\ref{subsec:glm.formula}, we provide more formulae including $\nu'(\eta)$'s, which are needed for the ForLion algorithm (see also Section~\ref{sec:GLM_sensitivity_derivative} in the Supplementary Material).

\par

\section{First-order derivative of sensitivity function for GLMs}\label{sec:GLM_sensitivity_derivative}

In this section, similar as in Appendix~\ref{sec:MLM_first_order}, to search for a new design point ${\mathbf x}^*$ in Step~$5^\circ$ of Algorithm~\ref{algo:ForLion_general}, we provide the first-order derivatives of $d({\mathbf x}, {\boldsymbol\xi}_t)$ for GLMs, given ${\mathbf x} = (x_1, \ldots, x_d)^T \in {\mathcal X}$.
Recall that the first $k$ factors are continuous with $x_i \in [a_i, b_i]$ and the last $d-k$ factors are discrete with $x_i \in I_i$, $1\leq k\leq d$. Denote ${\mathbf x}_{(1)} = (x_1, \ldots, x_k)^T$. 
To simplify the notation, we denote ${\mathbf A} = ({\mathbf X}_{{\boldsymbol\xi}_t}^T{\mathbf W}_{{\boldsymbol\xi}_t}{\mathbf X}_{{\boldsymbol\xi}_t})^{-1}$, which is a known positive definite matrix given ${\boldsymbol\xi}_t$ and $\boldsymbol\beta$. 
To perform the quasi-Newton method for Step~$5^\circ$ of Algorithm~\ref{algo:ForLion_general}, we need the partial gradient as follows
\begin{eqnarray*}
& & \frac{\partial d({\mathbf x}, {\boldsymbol\xi}_t)}{\partial {\mathbf x}_{(1)}} \\
&=& \nu'({\boldsymbol\beta}^T {\mathbf h}({\mathbf x}))\cdot {\mathbf h}({\mathbf x})^T {\mathbf A} {\mathbf h}({\mathbf x}) \cdot \left[\frac{\partial {\mathbf h}({\mathbf x})}{\partial {\mathbf x}_{(1)}^T}\right]^T {\boldsymbol\beta}\\
	&+& 2\cdot \nu({\boldsymbol\beta}^T {\mathbf h}({\mathbf x}))\cdot \left[\frac{\partial {\mathbf h}({\mathbf x})}{\partial {\mathbf x}_{(1)}^T}\right]^T {\mathbf A} {\mathbf h}({\mathbf x})
\end{eqnarray*}
where $\partial {\mathbf h}({\mathbf x})/\partial {\mathbf x}_{(1)}^T$ is a $p\times k$ matrix. Note that in general $d({\mathbf x}, {\boldsymbol\xi}_t)$ is not a convex function of ${\mathbf x}$ or ${\mathbf x}_{(1)}$~. Typically when $k\geq 2$, we need to try multiple initial  ${\mathbf x}_{(1)}$'s to bridge the gap between the local maximum and global maximum (see Remark~\ref{rem:global_maxima} in Section~\ref{sec:general_algorithm}).

To facilitate the readers, we provide the formulae of $\nu(\eta)$ and $\nu'(\eta)$ for commonly used GLMs in Section~\ref{subsec:glm.formula}.

\par

\section{GLMs with main-effects continuous factors}\label{sec:maineffects_supp}

In this section, we provide the formulae for a special class of GLMs with mixed factors, whose continuous factors are involved as main-effects only. 
That is, we assume that the predictors ${\mathbf h}({\mathbf x}) = (h_1({\mathbf x}), \ldots, h_p({\mathbf x}))^T$ satisfy
$h_i({\mathbf x}) = x_i$
for $i=1, \ldots, k$ and $h_{k+1}({\mathbf x}), \ldots, h_p({\mathbf x})$ do not depend on the continuous factors $x_1, \ldots, x_k$~. In other words, the linear predictor of the GLM model takes the form of
\begin{eqnarray}
\eta &=& \beta_1 x_1 + \cdots + \beta_k x_k + \beta_{k+1} h_{k+1}({\mathbf x}_{(2)})\nonumber\\
& & \ + \cdots + \beta_p h_p ({\mathbf x}_{(2)})
\label{eq:maineffects}
\end{eqnarray}
where ${\mathbf x}_{(2)} = (x_{k+1}, \ldots, x_d)^T$. Note that $h_{k+1}({\mathbf x}_{(2)})$ could be $1$ and then $\beta_{k+1}$ represents the intercept.

Recall that ${\mathbf x} = ({\mathbf x}_{(1)}^T, {\mathbf x}_{(2)}^T)^T$ and ${\mathbf x}_{(1)} = (x_1, \ldots, x_k)^T$. We denote $\boldsymbol\beta_{(1)} = (\beta_1, \ldots, \beta_k)^T$, $\boldsymbol\gamma_{(1)} = (\gamma_1, \ldots, \gamma_k)^T$, and ${\mathbf A}_{11}$ as the $k$th leading principle submatrix of ${\mathbf A} = ({\mathbf X}_{{\boldsymbol\xi}_t}^T{\mathbf W}_{{\boldsymbol\xi}_t}{\mathbf X}_{{\boldsymbol\xi}_t})^{-1}$, that is, the $k\times k$ submatrix of ${\mathbf A}$ formed by the first $k$ rows and the first $k$ columns. 

In this case, the $p\times k$ matrix $\partial {\mathbf h}({\mathbf x})/\partial {\mathbf x}_{(1)}^T = ({\mathbf I}_k\ {\mathbf 0})^T$, the $k\times k$ matrix $\partial^2 h_i({\mathbf x})/(\partial {\mathbf x}_{(1)} \partial {\mathbf x}_{(1)}^T)$ $ = {\mathbf 0}$ for $i=1, \ldots, p$, and
\[
\left[\frac{\partial {\mathbf h}({\mathbf x})}{\partial {\mathbf x}_{(1)}^T}\right]^T {\boldsymbol\beta} = \boldsymbol{\beta}_{(1)}, \>\>
\left[\frac{\partial {\mathbf h}({\mathbf x})}{\partial {\mathbf x}_{(1)}^T}\right]^T {\mathbf A} {\mathbf h}({\mathbf x}) = \boldsymbol{\gamma}_{(1)}
\]
Therefore, the partial gradient in this case is
\begin{eqnarray*}
\frac{\partial d({\mathbf x}, {\boldsymbol\xi}_t)}{\partial {\mathbf x}_{(1)}}  &=& \nu'({\boldsymbol\beta}^T {\mathbf h}({\mathbf x}))\cdot {\mathbf h}({\mathbf x})^T {\mathbf A} {\mathbf h}({\mathbf x}) \cdot {\boldsymbol\beta}_{(1)}\\
&+& 2\cdot \nu({\boldsymbol\beta}^T {\mathbf h}({\mathbf x}))\cdot \boldsymbol{\gamma}_{(1)}
\end{eqnarray*}	

The formulae in this section are useful for examples in Section~\ref{sec:maineffects}.

\par

\section{Electrostatic discharge example supplementary} \label{sec:mergedistance_delta}

In this section, we provide more details about Example~\ref{ex:ESD} on the electrostatic discharge experiment. 

Table~\ref{tab:ESD} shows the $14$-points design $\boldsymbol{\xi}_*$ obtained from the ForLion algorithm and the $13$-points design $\boldsymbol{\xi}_o$ from the $d$-QPSO algorithm of \cite{lukemire2018} (see Section~\ref{sec:maineffects}). They are fairly close to each other. As mentioned in Example~\ref{ex:ESD}, ForLion's $\boldsymbol{\xi}_*$ is slightly more efficient.

\begin{table*}[hbt!]
\centering
		\caption{Designs Obtained for Example~\ref{ex:ESD}}\label{tab:ESD}
\footnotesize
\begin{tabular}{|r|rrrrrr|rrrrrr|}\hline
		& \multicolumn{6}{c|}{d-QPSO\cite{lukemire2018}'s ${\boldsymbol\xi}_o$} & \multicolumn{6}{c|}{ForLion's ${\boldsymbol\xi}_*$}\\ \hline
$i$ & A & B & ESD & Pulse & Volt. & $w_i (\%)$ &  A & B & ESD & Pulse & Volt. & $w_i (\%)$\\ \hline
1  & -1  & -1  & -1  & -1  & 25.00  & 7.46  & -1  & -1  & -1  & -1  & 25.00  & 7.49 \\
2  & -1  & -1  & -1  & -1  & 28.04  & 1.80  & -1  & -1  & -1  & -1  & 27.55  & 1.56 \\
3  & -1  & -1  & -1  & 1  & 25.00  & 2.49  & -1  & -1  & -1  & 1  & 25.00  & 3.66 \\
4  & -1  & -1  & -1  & 1  & 27.85  & 7.74  & -1  & -1  & -1  & 1  & 28.69  & 7.22 \\
5  & -1  & -1  & 1  & -1  & 25.00  & 11.65  & -1  & -1  & 1  & -1  & 25.00  & 11.65 \\
6  & -1  & -1  & 1  & 1  & 25.00  & 8.58  & -1  & -1  & 1  & 1  & 25.00  & 8.54 \\
7  & -1  & 1  & -1  & -1  & 25.00  & 9.20  & -1  & 1  & -1  & -1  & 25.00  & 8.95 \\
8  & --  & --  & --  & --  & --  & --  & -1  & 1  & -1  & -1  & 29.06  & 0.42 \\
9  & -1  & 1  & -1  & 1  & 25.00  & 10.00  & -1  & 1  & -1  & 1  & 25.00  & 10.08 \\
10  & -1  & 1  & 1  & -1  & 25.00  & 3.80  & -1  & 1  & 1  & -1  & 25.00  & 3.41 \\
11  & -1  & 1  & 1  & -1  & 32.93  & 13.43  & -1  & 1  & 1  & -1  & 32.78  & 13.13 \\
12  & -1  & 1  & 1  & 1  & 25.00  & 9.20  & -1  & 1  & 1  & 1  & 25.00  & 9.23 \\
13  & 1  & -1  & 1  & -1  & 25.00  & 1.23  & 1  & -1  & 1  & -1  & 25.00  & 1.36 \\
14  & 1  & 1  & 1  & -1  & 25.00  & 13.40  & 1  & 1  & 1  & -1  & 25.00  & 13.31 \\
\hline
\end{tabular}
\end{table*}

\begin{remark}\label{rem:point8}
{\bf Removal of negligible design points:}\quad {\rm
In design $\boldsymbol{\xi}_*$ found by the ForLion algorithm, the $8$th design point with $w_8 = 0.42\%$ does not have a match in design $\boldsymbol{\xi}_o$. If we remove the $8$th design point and adjust $w_i$ to $w_i/(1-w_8)$, for $i\neq 8$, the resulting design, denoted by $\boldsymbol{\xi}_*'$, contains the same number of design points as in $\boldsymbol{\xi}_o$, but is slightly better than $\boldsymbol{\xi}_o$, with a relative efficiency $[f(\boldsymbol\xi_*')/f(\boldsymbol\xi_o)]^{1/7} = 100.08\%$. In practice, one can always remove negligible design points after checking the relative efficiency.
}\hfill{$\Box$}
\end{remark}

\begin{figure*}[t!] 
    \centering
    \begin{minipage}[t]{0.45\textwidth}
        \centering
        \includegraphics[height=2.4in]{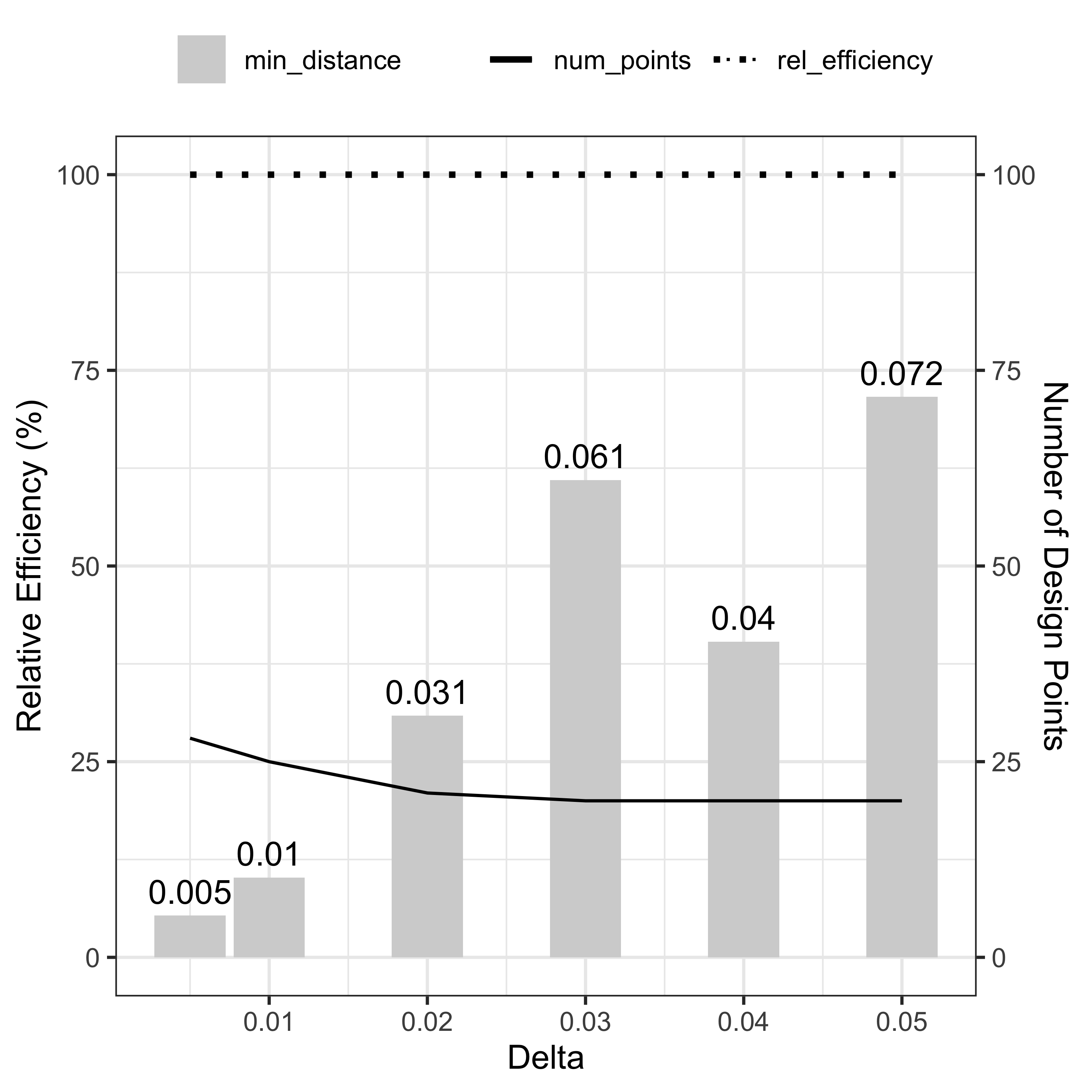}

        (a) Number of design points, minimum distance among design points, and relative efficiency w.r.t. the design with $\delta=0.005$
    \end{minipage}%
    ~ 
    \begin{minipage}[t]{0.45\textwidth}
        \centering
        \includegraphics[height=2.15in]{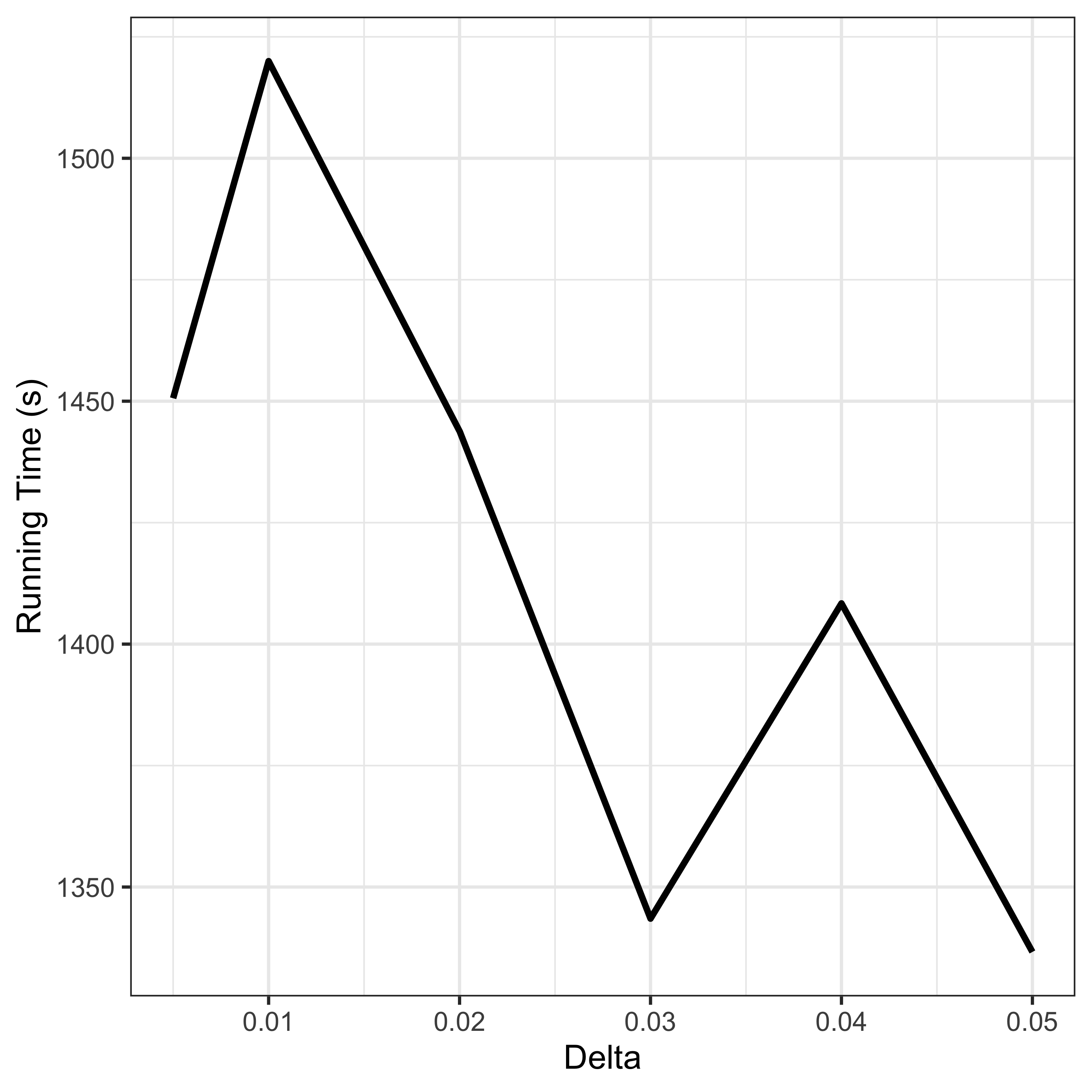}

        (b) Running time
    \end{minipage}
    \caption{Simulation results of ForLion algorithm with merging threshold $\delta \in\{ 0.05, 0.01, 0.02, 0.03, 0.04, 0.05\}$ while fixing $\epsilon = 10^{-8}$}\label{fig:glm_ESD_delta}
\end{figure*}

\begin{figure*}[t!] 
    \centering
    \begin{minipage}[t]{0.45\textwidth}
        \centering
        \includegraphics[height=2.15in]{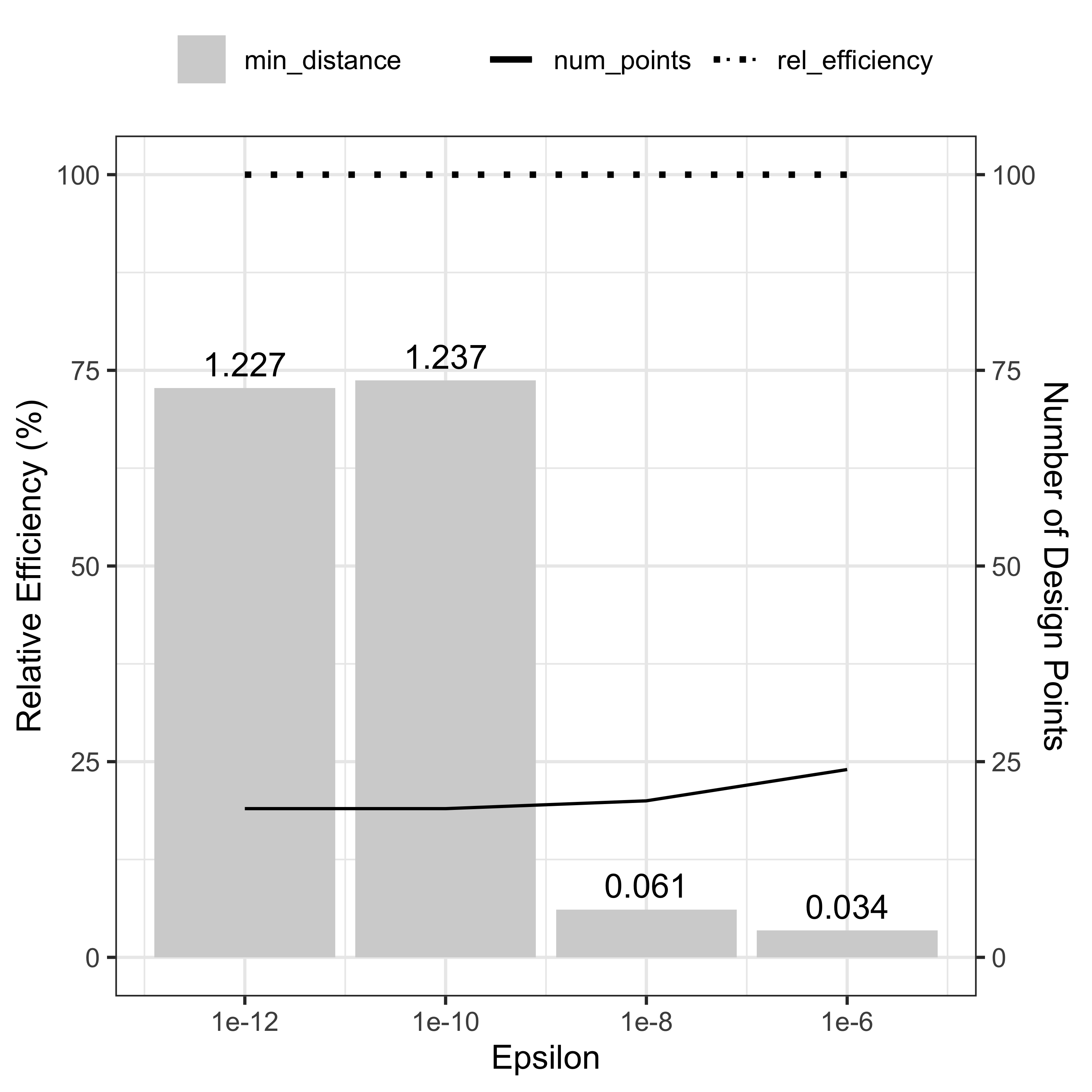}

        (a) Number of design points, minimum distance among design points, and relative efficiency w.r.t. the design with $\epsilon=10^{-8}$
    \end{minipage}%
    ~ 
    \begin{minipage}[t]{0.45\textwidth}
        \centering
        \includegraphics[height=2.15in]{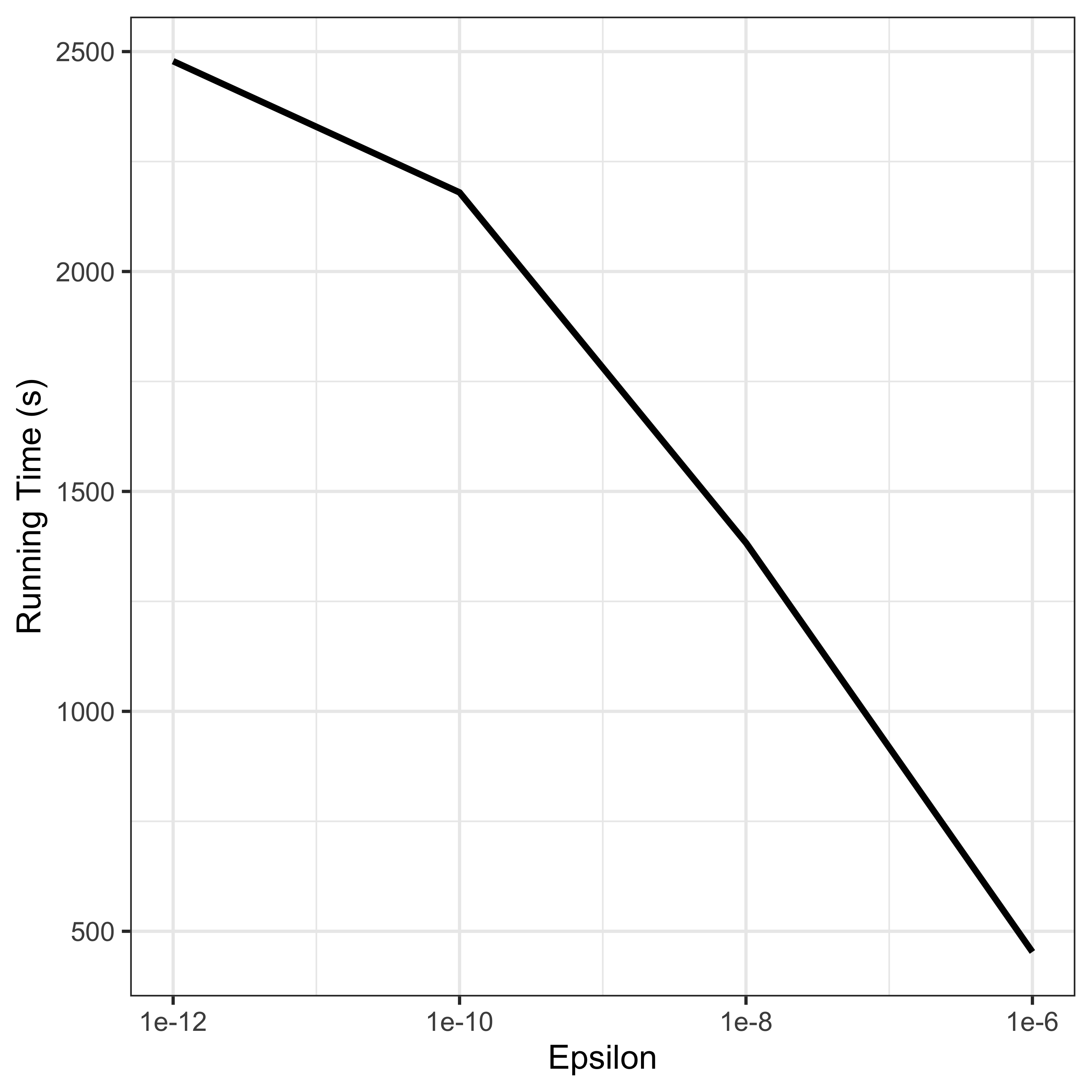}

        (b) Running time
    \end{minipage}
    \caption{Simulation results of ForLion algorithm with convergence threshold $\epsilon \in\{ 10^{-12}, 10^{-10}, 10^{-8}, 10^{-6}\}$ while fixing $\delta = 0.03$}\label{fig:glm_ESD_epsilon}
\end{figure*}

We also conduct a simulation study on Example~\ref{ex:ESD} to  investigates the impact of convergence threshold $\epsilon$ and merging threshold $\delta$ in Step~$0^\circ$ of Algorithm~\ref{algo:ForLion_general}. We assume that $\boldsymbol\beta = (\beta_0, \beta_1, \beta_2, \beta_3, \beta_4, \beta_5, \beta_{34})^T = (-7.84, 1.14, -0.21, 0, 0.38, 0.27, 0.45)^T$ are the true parameter values. The ForLion algorithm is employed to find D-optimal designs across varying thresholds $\delta \in \{0.005, 0.01, 0.02,$ $ 0.03, 0.04, 0.05\}$ and $\epsilon \in \{10^{-12}, 10^{-10}, 10^{-8}, 10^{-6}\}$. The results are shown in Tables~\ref{tab:esd_delta_epsilon_1}\nobreakdash--\ref{tab:esd_delta_epsilon_4} with relative efficiency defined as 
\[
(| \mathbf F(\cdot)|/| \mathbf F(\boldsymbol \xi_{\delta=0.03, \epsilon = 10^{-8}})|)^{1/7}
\]
where the design $\boldsymbol \xi_{\delta=0.03, \epsilon = 10^{-8}}$ obtained with tuning parameters $\delta=0.03$ and $\epsilon = 10^{-8}$ is used as a baseline for comparison purpose. Additionally, Figures~\ref{fig:glm_ESD_delta} and \ref{fig:glm_ESD_epsilon} illustrate the effects of merging threshold $\delta$ while fixing $\epsilon = 10^{-8}$ and the effects of convergence threshold $\epsilon$ while fixing  $\delta=0.03$.

From Figure~\ref{fig:glm_ESD_delta}, we can see that: (1) The number of design points decreases as we increase the merging threshold $\delta$, from $29$ when $\delta=0.005$ to $21$ when $\delta = 0.02$, and it then stabilizes at $20$ when $\delta \geq 0.03$; (2) the relative efficiency with respect to the design with $\delta=0.005$ remains around $100\%$; (3) the minimum distance among design points increases along with $\delta$; (4) the running time roughly shows a decreasing pattern with the minimum occurs at $\delta = 0.03$ or $\delta = 0.05$. Overall we choose $\delta=0.03$ for the simulation study described in Example~\ref{ex:ESD}.

From Figure~\ref{fig:glm_ESD_epsilon}, we can see that: Along with the increase of the convergence threshold $\epsilon$, (1) the number of design points increases from 20 to 24; (2) the relative efficiency remains roughly the same, around $100\%$; (3) the minimum distance between design points overall decreases; and (4) the running time drops. After all, we recommend $\epsilon = 10^{-8}$, which achieves a trade-off between the number of design points and the computational time cost in Example~\ref{ex:ESD}.

In conclusion, increasing the merging threshold $\delta$ typically results in fewer design points and reduced running time. Conversely, lowering the convergence threshold $\epsilon$ tends to yield fewer design points but at the cost of increased running time. Despite these variations, the efficiency of the designs remains robust across different combinations of $\delta$ and $\epsilon$. Selecting a good combination of these thresholds is crucial to the best design obtained. Our chosen tuning parameters of $\delta = 0.03$ and $\epsilon = 10^{-8}$ for Example~\ref{ex:ESD} are validated by the results.

\begin{table*}[hbt!]
\footnotesize
\centering
\caption{Relative efficiencies of designs w.r.t. design obtained with $\delta=0.03$ and $\epsilon = 10^{-8}$}\label{tab:esd_delta_epsilon_1}
\begin{tabular}{c|cccc}
\hline
 \backslashbox{merging $\delta$}{convergence $\epsilon$}& $10^{-12}$ & $10^{-10}$ & $10^{-8}$ & $10^{-6}$ \\\hline
 0.005 & 1.0000006 & 1.000001  & 1.0000003 & 0.9999866\\
 0.01 & 1.0000005  & 1.000001 &  1.0000004 &  0.9999879\\
 0.02 & 1.0000004 & 1.000000  & 1.0000003 &  0.9999910 \\
 0.03 & 1.0000004  & 1.000000 &  1.0000000 &  0.9999933\\
 0.04 & 1.0000000  & 1.000000  & 0.9999997 & 0.9999866\\
 0.05 &  0.9999995 & 1.000000  & 0.9999993 & 0.9999837 \\\hline
\end{tabular}
\normalsize
\end{table*}

\begin{table*}[hbt!]
\footnotesize
\centering
\caption{Number of design points in designs obtained for Example 3}\label{tab:esd_delta_epsilon_2}
\begin{tabular}{c|cccc}
\hline
 \backslashbox{merging $\delta$}{convergence $\epsilon$}& $10^{-12}$ & $10^{-10}$ & $10^{-8}$ & $10^{-6}$ \\\hline
 0.005 & 22 & 22 & 28 & 31\\
 0.01 & 21  & 19 & 25 & 26 \\
 0.02 & 20  & 21 & 21 & 25 \\
 0.03 & 19  & 19 & 20 & 24 \\
 0.04 & 20  & 19 & 20 & 25\\
 0.05 & 19  & 19 & 20 & 25 \\\hline
\end{tabular}
\normalsize
\end{table*}

\begin{table*}[hbt!]
\footnotesize
\centering
\caption{Minimum distance between design points  in designs obtained for Example 3}\label{tab:esd_delta_epsilon_3}
\begin{tabular}{c|cccc}
\hline
 \backslashbox{merging $\delta$}{convergence $\epsilon$}& $10^{-12}$ & $10^{-10}$ & $10^{-8}$ & $10^{-6}$ \\\hline
 0.005 & 0.0053 & 0.0055 & 0.0053 & 0.0062\\
 0.01 & 0.0151 & 1.2303  & 0.0102 & 0.0203\\
 0.02 & 0.0272  & 0.0263 & 0.0309 & 0.0253\\
 0.03 & 1.2272 & 1.2370 & 0.0610 & 0.0344 \\
 0.04 & 0.0689 & 1.2370 & 0.0403 & 0.0437\\
 0.05 & 1.2272 & 1.2370 & 0.0716 & 0.0551\\\hline
\end{tabular}
\normalsize
\end{table*}

\begin{table*}[hbt!]
\footnotesize
\centering
\caption{Runing time (s) for obtaining designs for Example 3}\label{tab:esd_delta_epsilon_4}
\begin{tabular}{c|cccc}
\hline
 \backslashbox{merging $\delta$}{convergence $\epsilon$}& $10^{-12}$ & $10^{-10}$ & $10^{-8}$ & $10^{-6}$ \\\hline
 0.005 & 2930.42 & 2620.22 & 1487.95 &  506.33\\
 0.01 & 2683.06  & 2210.02 & 1567.64 & 477.56\\
 0.02 & 2577.70 & 2436.48 & 1483.85 & 471.00\\
 0.03 & 2478.23  & 2179.77 & 1383.12 & 453.02\\
 0.04 & 2530.43  & 2178.93 &1447.19 & 415.60\\
 0.05 & 2404.10 &  2180.07 & 1376.50 & 412.64 \\\hline
\end{tabular}
\normalsize
\end{table*}

\section{Assumptions needed for Theorem~\ref{thm:doptimality}}\label{subsec:assumptions}

In this section, we list the assumptions needed for Theorem~\ref{thm:doptimality} for readers' reference (see also Section~2.4 of \cite{fedorov2014}):

\begin{itemize}
\item[(A1)] The design space ${\mathcal X}$ is compact.
\item[(A2)] The information matrix $\mathbf F_{\mathbf x}$ is continuous with respect to $\mathbf x \in {\mathcal X} $.
\item[(B1)] The optimality criterion $\Psi(\mathbf F)$ is convex, where ${\mathbf F}$ is the information matrix of some design.
\item[(B2)] $\Psi(\mathbf F)$ is monotonically nonincreasing with respect to ${\mathbf F}$.
\item[(B3)] There exists a real number $q$, such that, $\Xi(q)=\{\mathbf \xi: \Psi(\mathbf F(\boldsymbol \xi)) \le q < \infty\}$ is non-empty.
\item[(B4)] Let $\boldsymbol \xi$ and $\bar{\boldsymbol \xi}$ be two designs. If $\boldsymbol \xi \in \Xi(q)$, then 
\begin{eqnarray*}
    & &\Psi\left((1-\alpha)\mathbf F(\boldsymbol \xi) + \alpha \mathbf F(\bar{\boldsymbol \xi})\right)\\
    &=& \Psi\left(\mathbf F(\boldsymbol \xi)\right) + \alpha \int_{\mathcal X} \psi(\mathbf x, \boldsymbol \xi)\bar{\boldsymbol \xi}(d\mathbf x) + o(\alpha |\boldsymbol\xi, \bar{\boldsymbol \xi})
\end{eqnarray*} 
for some function $\psi$ of $\mathbf x$ and $\boldsymbol \xi$,  where $\lim_{\alpha \rightarrow 0^+} o(\alpha |\boldsymbol\xi, \bar{\boldsymbol \xi})/\alpha=0$.
\end{itemize}

For D-optimality, $\Psi({\mathbf F}) = -\log|\mathbf F|$ and $\psi(\mathbf x, \boldsymbol \xi) = p - {\rm tr}\left(\mathbf F^{-1}(\boldsymbol \xi)\mathbf F_{\mathbf x}\right)$. 

For all the design spaces and statistical models discussed in this paper, (A1) and (A2) are satisfied. According to Sections~2.4 and 2.5 in \cite{fedorov2014}, D-optimality satisfies (B1)-(B4).

\section{Proofs}\label{subsec:proofs}

\noindent
{\bf Proof of Theorem~\ref{thm:doptimality}:} 
As a direct conclusion of Theorem~2.2 of \cite{fedorov2014}, under regularity conditions, there exists a D-optimal design that contains no more than $p(p+1)/2$ support points, which are the design points with positive weights.

According to Theorem~2.2 of \cite{fedorov2014}, a design $\boldsymbol{\xi}$ is optimal if and only if $\min_{{\mathbf x} \in {\mathcal X}} \psi({\mathbf x}, \boldsymbol{\xi}) \geq 0$. Under D-criterion, it is equivalent to $\max_{{\mathbf x} \in {\mathcal X}} d({\mathbf x}, \boldsymbol{\xi}) \leq p$ in our notations. If a design $\boldsymbol{\xi}$ is reported by Algorithm~\ref{algo:ForLion_general}, it must satisfy $\max_{{\mathbf x} \in {\mathcal X}} d({\mathbf x}, \boldsymbol{\xi}) \leq p$ and thus be D-optimal.

\hfill{$\Box$}

\medskip
\noindent
{\bf Proof of Theorem~\ref{thm:F_x}:} 
It is a direct conclusion of  Corollary~3.1 of \cite{bu2020}, ${\mathbf F}_{\mathbf x} = {\mathbf X}_{\mathbf x}^T {\mathbf U}_{\mathbf x} {\mathbf X}_{\mathbf x}$, where ${\mathbf U}_{\mathbf x} = (u_{st}^{\mathbf x})_{s,t=1,\ldots, J}$.
\hfill{$\Box$}

\begin{lemma}\label{lem:trace_block_matrices}
Let 
\[
{\mathbf A} = \left[\begin{array}{ccc}
{\mathbf A}_{11} & \cdots & {\mathbf A}_{1J}\\
\vdots & \ddots & \vdots\\
{\mathbf A}_{J1} & \cdots & {\mathbf A}_{JJ}
\end{array}\right], \>\>\>
{\mathbf B} = \left[\begin{array}{ccc}
{\mathbf B}_{11} & \cdots & {\mathbf B}_{1J}\\
\vdots & \ddots & \vdots\\
{\mathbf B}_{J1} & \cdots & {\mathbf B}_{JJ}
\end{array}\right]
\]
be two $n\times n$ symmetric matrices with submatrices ${\mathbf A}_{ij}, {\mathbf B}_{ij} \in \mathbb{R}^{n_i\times n_j}$, $i,j=1, \ldots, J$, $\sum_{i=1}^J n_i = n$. Then 
\begin{eqnarray*}
& & {\rm tr}({\mathbf A} {\mathbf B}) \ = \ \sum_{i=1}^J \sum_{j=1}^J {\rm tr}({\mathbf A}_{ij} {\mathbf B}_{ij}^T)\\ &=& \sum_{j=1}^J {\rm tr}({\mathbf A}_{jj} {\mathbf B}_{jj})
+ 2 \sum_{i=1}^{J-1} \sum_{j=i+1}^J {\rm tr}({\mathbf A}_{ij} {\mathbf B}_{ij}^T)
\end{eqnarray*}
\hfill{$\Box$}
\end{lemma}

\medskip
\noindent
{\bf Proof of Lemma~\ref{lem:trace_block_matrices}:} It is a direct conclusion of the definition of trace. Please see, for example, Formulae~4.13 (a) and (b) in \cite{seber2008}.
\hfill{$\Box$}

\medskip
\noindent
{\bf Proof of Theorem~\ref{thm:generalequivalence_mlm}:} According to Theorem~2.3 and Table~2.1 in \cite{fedorov2014}, $\boldsymbol\xi$ is D-optimal if and only if $\max_{{\mathbf x} \in {\mathcal X}} d({\mathbf x}, {\boldsymbol\xi}) \leq p$, where $d({\mathbf x}, \boldsymbol\xi) = {\rm tr}({\mathbf F}(\boldsymbol\xi)^{-1} {\mathbf F}_{\mathbf x})$. Equation~\eqref{eq:d(x,xi)_GLM} can be obtained by applying Lemma~\ref{lem:trace_block_matrices} and Theorem~\ref{thm:F_x}, as well as 
\begin{eqnarray*}
{\rm tr}({\mathbf C}_{ij} u_{ij}^{\mathbf x} {\mathbf h}_i^{\mathbf x} ({\mathbf h}_j^{\mathbf x})^T) &=& u_{ij}^{\mathbf x} {\rm tr}(({\mathbf h}_j^{\mathbf x})^T {\mathbf C}_{ij}  {\mathbf h}_i^{\mathbf x})\\
&=& u_{ij}^{\mathbf x} ({\mathbf h}_j^{\mathbf x})^T {\mathbf C}_{ij}  {\mathbf h}_i^{\mathbf x}
\end{eqnarray*}
\hfill{$\Box$}

\medskip
\noindent
{\bf Proof of Theorem~\ref{thm:generalequivalence}:} For D-criterion, for example, see \citeauthor{fedorov2014} (2014, section~2.5), $\boldsymbol\xi$ is D-optimal if and only if
\begin{eqnarray*}
& &\max_{{\mathbf x} \in {\mathcal X}} {\rm tr}\left[({\mathbf X}_{\boldsymbol\xi}^T{\mathbf W}_{\boldsymbol\xi}{\mathbf X}_{\boldsymbol\xi})^{-1}\cdot \nu({\boldsymbol\beta}^T {\mathbf h}({\mathbf x})) {\mathbf h}({\mathbf x}) {\mathbf h}({\mathbf x})^T\right]\\
&=&\max_{{\mathbf x} \in {\mathcal X}} \nu({\boldsymbol\beta}^T {\mathbf h}({\mathbf x})) {\mathbf h}({\mathbf x})^T ({\mathbf X}_{\boldsymbol\xi}^T{\mathbf W}_{\boldsymbol\xi}{\mathbf X}_{\boldsymbol\xi})^{-1} {\mathbf h}({\mathbf x}) \leq  p
\end{eqnarray*}
\hfill{$\Box$}

\medskip
\noindent
{\bf Proof of Theorem~\ref{lemma:alphat}:}
We are given the design at the $t$th iteration ${\boldsymbol\xi}_t = \{({\mathbf x}^{(t)}_i, w^{(t)}_i), i=1, \ldots, m_t\}$ and the new design point ${\mathbf x}^*$. We define a function of the allocation for the $(t+1)$th iteration $f^{(t)} (w_1, \ldots, w_{m_t+1}) = f(\{({\mathbf x}^{(t)}_1, w_1), \ldots, ({\mathbf x}^{(t)}_{m_t}, w_{m_t}), ({\mathbf x}^*, w_{m_t+1})\})$.

Following the notations for the lift-one algorithm (see (1) in \cite{ym2015}), we define $f^{(t)}_{m_t+1}(x) = f^{(t)}((1-x)w^{(t)}_1, \ldots, (1-x) w^{(t)}_{m_t}, x)$ for $0\leq x\leq 1$. Then $f({\boldsymbol\xi}_t) = f^{(t)}(w^{(t)}_1, \ldots, w^{(t)}_{m_t}, 0) = f^{(t)}_{m_t+1}(0)$.
According to Lemma~4.1 in \cite{ym2015}, $f^{(t)}_{m_t+1}(x) = a x(1-x)^{p-1} + b(1-x)^p$ with $b = f(\boldsymbol\xi_t) \stackrel{\triangle}{=} b_t$ and $a = f^{(t)}_{m_t+1}(1/2) \cdot 2^p - b \stackrel{\triangle}{=} d_t \cdot 2^p - b_t$. The conclusion then can be drawn from Lemma~4.2 in \cite{ym2015}.
\hfill{$\Box$}

\medskip
\noindent
{\bf Proof of Theorem~\ref{thm:minimallysupported}:}
${\mathbf X}_{\boldsymbol\xi} = ({\mathbf h}({\mathbf x}_1),$ $\ldots,$ ${\mathbf h}({\mathbf x}_m))^T$ is $m\times p$, ${\mathbf W}_{\boldsymbol\xi} = {\rm diag}\{w_1 \nu({\boldsymbol\beta}^T {\mathbf h}({\mathbf x}_1)),$ $\ldots,$ $w_m \nu({\boldsymbol\beta}^T {\mathbf h}({\mathbf x}_m))\}$ is $m\times m$, and ${\mathbf X}_{\boldsymbol\xi}^T {\mathbf W}_{\boldsymbol\xi} {\mathbf X}_{\boldsymbol\xi}$ is $p\times p$. Then  $f({\boldsymbol\xi}) > 0$ implies $p = {\rm rank}({\mathbf X}_{\boldsymbol\xi}^T {\mathbf W}_{\boldsymbol\xi} {\mathbf X}_{\boldsymbol\xi}) \leq \min\{m, {\rm rank}({\mathbf X}_{\boldsymbol\xi})\}$. That is, ${\rm rank}({\mathbf X}_{\boldsymbol\xi}) = p \leq m$. If furthermore, $\nu({\boldsymbol\beta}^T {\mathbf h}({\mathbf x}_i)) > 0$ for all $i=1, \ldots, m$, that is, ${\mathbf W}_{\boldsymbol\xi}$ is of full rank, then ${\rm rank}({\mathbf X}_{\boldsymbol\xi}^T {\mathbf W}_{\boldsymbol\xi} {\mathbf X}_{\boldsymbol\xi}) = {\rm rank}({\mathbf X}_{\boldsymbol\xi})$. Therefore, $f({\boldsymbol\xi}) > 0$ if and only if ${\rm rank}({\mathbf X}_{\boldsymbol\xi}) = p$.
\hfill{$\Box$}

\medskip
\noindent
{\bf Proof of Theorem~\ref{thm:minimalluniform}:}
In this case, $m=p$, ${\mathbf X}_{\boldsymbol\xi}$ is $p\times p$, and
$f({\boldsymbol\xi}) = |{\mathbf X}_{\boldsymbol\xi}^T {\mathbf W}_{\boldsymbol\xi} {\mathbf X}_{\boldsymbol\xi}|
= |{\mathbf X}_{\boldsymbol\xi}|^2 \cdot |{\mathbf W}_{\boldsymbol\xi}|
= |{\mathbf X}_{\boldsymbol\xi}|^2 \prod_{i=1}^p \nu({\boldsymbol\beta}^T {\mathbf h}({\mathbf x}_i)) \cdot \prod_{i=1}^p w_i$.
Since $\prod_{i=1}^p w_i \leq p^{-p}$ with equality attaining at $w_i = p^{-1}$, then $\boldsymbol\xi$ is D-optimal only if $w_i = p^{-1}$, $i=1, \ldots, p$.

\end{document}